\documentclass[aps,prd,reprint,superscriptaddress,preprintnumbers,showpacs,nofootinbib]{revtex4-1}

\usepackage[colorlinks, pdfborder={0 0 0}, plainpages=false]{hyperref}
\usepackage[utf8]{inputenc}
\usepackage{graphicx}
\usepackage{amsmath}
\usepackage{amssymb}
\usepackage{acronym}
\usepackage{amsfonts,pifont}
\usepackage{txfonts}
\usepackage[normalem]{ulem}

\newcommand{\cmark}{\ding{51}}%
\newcommand{\xmark}{\ding{55}}%

\newcommand{\AEIHannover}{Max Planck  Institute for Gravitational Physics
(Albert Einstein Institute), Callinstr.~38, 30167 Hannover, Germany}
\newcommand{\UniHannover}{Leibniz Universit\"at Hannover, D-30167 Hannover, Germany}

\newcommand{\cardiff}{School of Physics and Astronomy, Cardiff University, The
Parade, Cardiff, CF24 3AA, United Kingdom}

\newcommand{\CCA}{Center for Computational Astrophysics, Flatiron Institute, 162 5th Ave, New York, NY 10010}

\newcommand{\rome}{Dipartimento di Fisica, Universit\`a di Roma ``Sapienza'', Piazzale A. Moro 5, I-00185 Roma, Italy}

\newacro{BH}{Black Hole}
\newacro{BBH}{Binary Black Hole}
\newacro{NR}{Numerical Relativity}
\newacro{GW}{Gravitational Wave}
\newacro{IMR}{Inspiral-Merger-Ringdown}
\newacro{NS}{Neutron Star}
\newacro{BNS}{Binary Neutron Star}
\newacro{LSC}{LIGO Scientific Collaboration}
\newacro{LVC}{LIGO-Virgo Collaboration}
\newacro{LIGO}{Laser Interferometric Gravitational-Wave Observatory}
\newacro{aLIGO}{Advanced LIGO}
\newacro{Adv}{Advanced Virgo}
\newacro{EM}{electromagnetic}
\newacro{PN}{Post Newtonian}
\newacro{EOB}{Effective-One-Body}
\newacro{MSA}{multiple scale analysis}
\newacro{SUA}{shifted uniform asymptotics}
\newacro{SPA}{stationary phase approximation}
\newacro{SNR}{signal-to-noise ratio}
\newacro{PSD}{power spectral density}
\newacro{ROQ}{reduced order quadrature}

\newcommand{\chieff}{\chi_{\rm{eff}}}

\begin{document}

\title{ Including higher order multipoles in gravitational-wave models for precessing binary black holes }

\author{Sebastian Khan}
\affiliation\cardiff
\affiliation\AEIHannover
\affiliation\UniHannover

\author{Frank Ohme}
\affiliation\AEIHannover
\affiliation\UniHannover

\author{Katerina Chatziioannou}
\affiliation\CCA

\author{Mark Hannam}
\affiliation\cardiff
\affiliation\rome


\date{\today}

\begin{abstract}

Estimates of the source parameters of gravitational-wave (GW) events produced by
compact binary mergers rely on theoretical
models for the GW signal.
We present the first frequency-domain model for inspiral, merger and ringdown
of the GW signal from
precessing binary-black-hole systems that also includes multipoles
beyond the leading-order quadrupole.
Our model, {\tt PhenomPv3HM}, is a combination of the
higher-multipole non-precessing model {\tt PhenomHM} and the spin-precessing model {\tt PhenomPv3}
that includes two-spin precession via a dynamical rotation of the
GW multipoles.
We validate the new model by comparing to a large
set of precessing numerical-relativity simulations
and find excellent agreement across the majority of the parameter
space they cover. For mass ratios $<5$ the mismatch
improves, on average, from $\sim6\%$ to $\sim 2\%$ compared to {\tt PhenomPv3} when
we include higher multipoles in the model.
However, we find mismatches $\sim8\%$ for a mass-ratio $6$ and highly
spinning simulation.
We quantify the statistical uncertainty in the recovery of binary
parameters by applying standard Bayesian parameter estimation methods
to simulated signals. We find that, while the primary black hole spin
parameters should be measurable even at moderate signal-to-noise ratios (SNRs)
$\sim 30$, the secondary spin requires much larger SNRs $\sim 200$.
We also quantify the systematic uncertainty expected by recovering
our simulated signals with different waveform models in which various
physical effects, such as the inclusion of higher modes and/or precession,
are omitted and find that even at the low SNR case ($\sim 17$)
the recovered parameters can be biased.
Finally, as a first application of the new model we have analysed the
binary black hole event GW170729. We find larger values for the
primary black hole mass of $58.25^{+11.73}_{-12.53} \, M_\odot$
(90\% credible interval).
The lower limit ($\sim 46 \, M_\odot$) is comparable to the proposed maximum black hole
mass predicted by different stellar evolution models due to the pulsation pair-instability supernova (PPISN) mechanism.
If we assume that the primary \ac{BH} in GW170729 formed through a PPISN
then out of the four PPISN models we considered only the model of
~\citet{Woosley_2017} is consistent with our mass measurements at the 90\% level.
\end{abstract}


\maketitle

\section{Introduction}
The second generation \ac{GW} detectors — Advanced LIGO~\cite{TheLIGOScientific:2014jea}
and Virgo~\cite{TheVirgo:2014hva} — have so far published observations of
11 compact binary mergers from the
first two observing runs~\cite{LIGOScientific:2018mvr}, including
one binary neutron star merger that was also observed across the electromagnetic
spectrum~\cite{GBM:2017lvd}.
The third observing run is currently underway, with further sensitivity improvements planned in the coming
years~\cite{Aasi:2013wya}.
\ac{GW} observations have already begun to constrain models of the formation and rates of stellar mass
compact binary mergers~\cite{LIGOScientific:2018jsj}, and to make strong-field tests of the general theory of
relativity~\cite{LIGOScientific:2019fpa}.

Models for the \ac{GW} signal, parametrized in terms of the properties of the system (such as masses, spins, and orientation), are compared with detector data to
infer the source properties of \ac{GW} events.
The \ac{GW} signal is commonly expressed in a multipole expansion where we
denote terms beyond the leading order quadrupole contribution as ``higher order multipoles''.
These higher order multipoles are typically much weaker than
the dominant quadrupolar multipole, but grow in relative strength
for systems that are more asymmetric in mass.
Past studies have shown that for events where the signal contains measurable
power in the higher multipoles,
parameter estimates can be biassed when using only a dominant-multipole model.
Conversely, it is also true that for some systems we are able to measure
the source parameters more accurately using a higher multipole model
~\cite{PhysRevD.92.022002,PhysRevLett.120.161102,Chatziioannou:2019dsz,Kalaghatgi:2019log,PhysRevLett.121.191102,2019arXiv191102693S}.

Another important physical effect is spin precession, where
couplings
between the orbital and spin angular
momenta can cause the orbital plane to precess
and thus cause modulations of the observed \ac{GW}~\cite{PhysRevD.49.6274,PhysRevD.52.821}.
In terms of the \ac{GW} multipoles, precession
mixes together different orders ($m$-multipoles) of the same degree ($\ell$-multipoles), complicating
a simple description of the
waveform~\cite{Schmidt:2010it,PhysRevD.84.124011,PhysRevD.85.084003,PhysRevD.86.104063,PhysRevD.88.024040,Boyle:2014ioa,PhysRevD.84.124002,Fairhurst:2019vut}.
By not taking into account precession and higher order multipoles
in our waveform models we may not be able to confidently
detect and accurately characterize signals where
these effects are important~\cite{Capano:2013raa,PhysRevD.96.124024,CalderonBustillo:2017skv,PhysRevD.94.024012,Harry:2017weg,PhysRevD.93.084019}.
These events are also likely to be very interesting astrophysically,
providing valuable information about \ac{BBH} formation mechanisms
and hence are events with high
scientific gain that we wish to model and measure accurately.

The field of waveform modeling has seen sustained development over almost two
decades and is currently thriving,
with improvements to current and development of novel methods
allowing for more accurate and efficient
models to be applied in data analysis pipelines
~\cite{PhysRevD.59.084006,Pan:2013rra,PhysRevD.62.064015,Pan:2011gk,
Taracchini:2013rva,Blanchet2014,Hannam:2013oca,Ajith:2007qp,
PhysRevD.82.064016,Khan:2015jqa,PhysRevLett.120.161102,
PhysRevD.98.084028,PhysRevD.95.044028,PhysRevD.95.024010,
PhysRevD.96.024058, PhysRevD.98.104052,Mehta:2019wxm,
Varma:2018mmi, Williams:2019vub,Setyawati:2018tqp,
PhysRevD.100.024059,PhysRevD.89.044021,Doctor:2017csx,Blackman:2017dfb,Varma:2019csw}.
In this work we take a step towards including as many important physical
effects as possible in waveform models, by constructing the most physically complete
phenomenological model to date.
We present a frequency-domain model for the \ac{GW} signal
from the inspiral, merger and ringdown of a \ac{BBH} system.
The \acp{BH} are allowed to precess and we also model the contribution
to the \ac{GW} signal from higher order multipoles.
This combines the progress made in two earlier models: a precessing-binary model
that includes accurate two-spin precession effects during the
inspiral~\cite{Chatziioannou:2017tdw,Chatziioannou:2016ezg}
({\tt PhenomPv3}~\cite{PhysRevD.100.024059}), and an approximate higher-multipole
aligned-spin model ({\tt PhenomHM}~\cite{PhysRevLett.120.161102}).

\begin{figure}[t!]
\includegraphics[width=0.98\columnwidth]{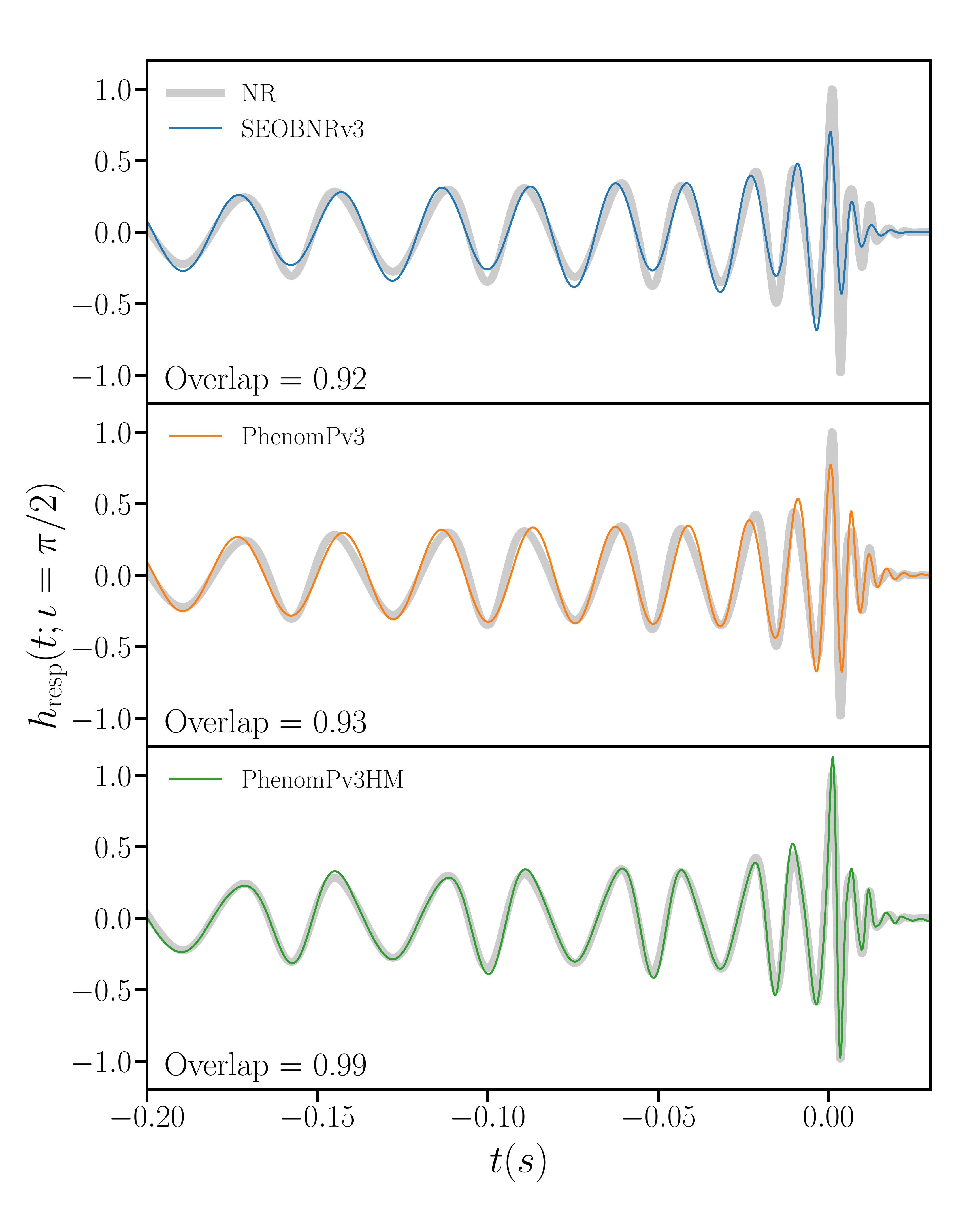}
 \caption{
 Comparison of the detector response strain $h(t)$ viewed at an inclination angle of $\pi/2$.
 Solid grey: \ac{NR} simulation (SXS:BBH:0058).
 A mass-ratio $q \equiv m_1/m_2 = 5$, precessing \ac{BBH}
 simulation with a dimensionless-spin magnitude of $\chi_1 = 0.5$ generated with a total mass of $80M_\odot$.
 The \ac{NR} signal contains all the $\ell,m$ modes up to and including $\ell=4$.
 Top panel, blue: Precessing model {\tt SEOBNRv3}~\cite{PhysRevD.89.084006}, with the $((2,\pm 2), (2,\pm 1))$
 modes in the co-precessing frame.
 Middle panel, orange: Precessing model {\tt PhenomPv3}~\cite{PhysRevD.100.024059} with only the $2,\pm 2$
 modes in the co-precessing frame.
 Bottom panel, green: Precessing model presented here, {\tt PhenomPv3HM}, with the
 $(\ell, |m|) = ((2,2), (2,1), (3,3), (3,2), (4,4), (4,3))$ modes in the co-precessing frame.
 The orientation-averaged mismatch $1-\overline{\mathcal{M}}$ (see Sec.~\ref{sec:matches})
 is $8\,\%$ for the top panel,
 $7\,\%$ for the middle and $1\,\%$ for the bottom.
 We only plot the last $\sim$7 \ac{GW} cycles for clarity but the behaviour is qualitatively
 the same throughout the 29-orbit inspiral ($\sim$60 \ac{GW} cycles).
 }
 \label{fig:TDComparison}
\end{figure}

Figure~\ref{fig:TDComparison} demonstrates the improved accuracy that is
achievable by our new model, {\tt PhenomPv3HM}, compared to other existing models
that include the effect of spin precession, but not higher order multipoles.
We compare the observed \ac{GW} signal
predicted by our new model against a high-mass-ratio, precessing
\ac{NR} simulation\footnote{The \ac{NR} waveform is SXS:BBH:0058 from the SXS
public catalogue~\cite{sxsonline}. It has a
mass-ratio of $q=5$ with spin on the larger \ac{BH}
directed in the orbital plane with a dimensionless spin magnitude of $\chi_1 = 0.5$.} (thick grey line).
We plot the \ac{GW} signal observed at an inclination
angle\footnote{Here we define the inclination as the angle between the orbital
angular momentum and the line of sight at the beginning of the waveform.}
of $\pi/2$ rad to emphasise the effect of precession.
We use all multipoles in the range $2 \leq \ell \leq 4$ when computing
the \ac{NR} \ac{GW} polarisations.

We compute the mismatch (defined in Section~\ref{sec:matches}) between
three different precessing waveform models and the \ac{NR} waveform, and average
over all possible orientations.
The top panel shows the optimal
waveform (in blue) when we use
{\tt SEOBNRv3}~\cite{PhysRevD.95.024010} and
the middle panel shows (in orange) the result when
we use {\tt PhenomPv3}.
In this context the optimal waveform maximises the overlap over
coalescence time, template phase and polarisation angle and the intrinsic
parameters are fixed to the values from the \ac{NR} simulation.
As shown in~\cite{PhysRevD.100.024059} {\tt SEOBNRv3} and {\tt PhenomPv3}
have overlaps of $\sim99\,\%$ and $\sim 98\,\%$ respectively to this \ac{NR}
waveform when only the $\ell=2$ multipoles are considered.
When we include higher order multipoles in the \ac{NR} waveform
we find the overlap drops to only $\sim 92\,\%$ and $\sim 93\,\%$ respectively.
This is an example where the exclusion of higher multipoles in template
models can lead to unacceptable losses in \ac{SNR}.
The bottom panel shows, in green, the best fitting {\tt PhenomPv3HM}
template.
We find remarkable agreement, even through the inspiral, merger
and ringdown stages. The overlap is now $99\,\%$ and the subtle modulation
visible is accurately captured by our model.
It is useful to point out here that, in {\tt PhenomPv3HM}, the higher multipole
and the precession elements of the model have not been
calibrated to \ac{NR} simulations, but when this is done
we expect the accuracy to improve further.

The rest of the paper is organized as follows. In Sec.~\ref{sec:method}
 we describe how our model is constructed.
 In Sec.~\ref{sec:matches} we present results where we have compared our
model against precessing \ac{NR} simulations including higher order mutlipoles
up to and including $\ell = 4$ to demonstrate its accuracy across
the parameter space where we have \ac{NR} simulations.
We have also performed a parameter estimation study to quantify the impact
on parameter recovery when using a model that includes both higher multipoles
and precession, the results of which are presented in section~\ref{sec:pe}.

Finally, in section~\ref{sec:gw170729} we have analysed data for
the GW170729 event, publicly available
at the Gravitational Wave Open Science Center
(GWOSC)~\cite{Vallisneri:2014vxa},
which has evidence for non-zero \ac{BH} spin~\cite{LIGOScientific:2018mvr}
and unequal masses~\cite{Chatziioannou:2019dsz}.

\section{Method}
\label{sec:method}

Our method to build a model for the \ac{GW} signal from precessing
\acp{BBH} is based upon the novel ideas of
Refs.~\cite{PhysRevD.84.024046, PhysRevD.84.124002, PhysRevD.84.124011},
where the \ac{GW} from precessing binaries
can be modelled as a dynamic rotation of non-precessing systems.
In  Refs.~\cite{Hannam:2013oca,Taracchini:2013rva,PhysRevD.89.084006} the authors
used these ideas to build the first precessing \ac{IMR} models.

Our goal is to derive frequency-domain expressions for the
\ac{GW} polarisations $\tilde{h}_{+/\times}(f)$ in terms of the
 multipoles $\tilde{h}_{\ell m}(f)$.
We start from the complex \ac{GW} quantity, $h = h_{+} - i h_{\times}$,
in the time-domain and decompose
this into spin weight $-2$ spherical harmonics,

\begin{equation}
\label{equ:htd}
  h(t,\vec{\lambda},\theta,\phi)  = \sum_{\ell \geqslant 2} \sum_{-\ell\leqslant m \leqslant \ell} h_{\ell,m}(t,\vec{\lambda}) {}_{-2}Y_{\ell,m}(\theta,\phi) \, .
\end{equation}

This is a function of the time $t$, the intrinsic source parameters (masses and spin angular momenta of the bodies)
 denoted by $\vec{\lambda}$, and the polar angles $\theta$ and $\phi$ of a coordinate system
whose $z-$axis is aligned with the total angular momentum $\vec{J}$ of the binary
at some reference frequency.
To approximate the precessing multipoles $h^{prec}_{\ell,m}(t)$ we
perform a dynamic rotation of the non-precessing multipoles
$h_{\ell,m}^{ \substack{non- \\ prec}}(t)$,

\begin{equation}
\label{equ:rot}
h^{prec}_{\ell,m}(t) = \sum_{-\ell\leqslant m' \leqslant \ell} h^{ \substack{non- \\ prec}  }_{\ell,m'}(t) \, D^{\ell}_{m',m}(\alpha(t), \beta(t), \epsilon(t)) \, .
\end{equation}

We define the Wigner D-matrix as
$D^{\ell}_{m',m}(\alpha, \beta, \epsilon) = e^{i m \alpha} d^{\ell}_{m',m}(-\beta) e^{-i m' \epsilon}$
and the Wigner \emph{d}-matrix is given in Ref.~\cite{Ajith:2007jx}.

Next we transform to the frequency domain using the stationary phase approximation~\cite{Droz:1999qx}
under the assumption that the precession angles modify the signal via a slowly
varying amplitude, giving us an expression for the frequency-domain
multipoles in terms of the co-precessing frame multipoles,

\begin{equation}
\label{equ:hlmprec}
\tilde{h}^{prec}_{\ell,m}(f) = \sum \limits_{-\leqslant \ell< m' \leqslant \ell} \tilde{h}^{ \substack{non- \\ prec}  }_{\ell,m'}(f) \, D^{\ell}_{m',m}(\alpha, \beta, \epsilon) \, .
\end{equation}

For brevity we omit the explicit dependence on frequency for the
the precession angles $(\alpha, \beta, \epsilon)$ but they
are evaluated at the stationary points $t(f) = 2\pi f/m'$
~\cite{PhysRevD.82.064016}.

The frequency-domain \ac{GW} polarisations $\tilde{h}_{+/\times}(f)$ are defined as the
Fourier transform (FT) of the real-valued \ac{GW} polarisations $h_{+/\times}(t)$,
which we write as,

\begin{eqnarray}
  \label{equ:hp_1}
  \tilde{h}_{+}(f) &=& {\rm{FT}}\left[ {\rm{Re}}(h(t)) \right] =  \frac{1}{2}\left(\tilde{h}(f) + \tilde{h}^{*}(-f)\right) \, , \\
  \label{equ:hc_1}
  \tilde{h}_{\times}(f) &=& {\rm{FT}}\left[ {\rm{Im}}(h(t)) \right] =  \frac{i}{2}\left(\tilde{h}(f) - \tilde{h}^{*}(-f)\right) \, .
\end{eqnarray}

To arrive at the final expression for the frequency-domain \ac{GW} polarisations
we substitute Eq.~(\ref{equ:hlmprec}) into Eqs.~(\ref{equ:hp_1}) and (\ref{equ:hc_1}),
assuming $f>0$ and symmetry through the orbital plane in the co-precessing
frame\footnote{This leads to the simplification $\tilde{h}_{\ell, m'}(f) = (-1)^{\ell} \tilde{h}^{*}_{\ell, -m'}(-f)$.}, leading to,

  \begin{align}
    \tilde{h}^{prec}_{+}(f) &= \frac{1}{2} \sum_{\ell \geqslant 2} \sum_{m'>0}
 \tilde{h}_{\ell,m'}^{ \substack{non- \\ prec}}(f) \sum_{m=-\ell}^{\ell} \left( A^{\ell}_{m',m} + (-1)^{\ell} A^{*\ell}_{-m',m} \right) \, ,
 \label{eq:hp}
  \end{align}
  \begin{align}
\tilde{h}^{prec}_{\times}(f) &= -\frac{i}{2} \sum_{\ell \geqslant 2} \sum_{m'>0}
 \tilde{h}_{\ell,m'}^{ \substack{non- \\ prec}}(f) \sum_{m=-\ell}^{\ell} \left( A^{\ell}_{m',m} - (-1)^{\ell} A^{*\ell}_{-m',m} \right) \, .
  \label{eq:hc}
  \end{align}

To shorten the expression we define the auxillary matrix
$A^{\ell}_{m',m} \equiv {}_{-2}Y_{\ell,m} D^{\ell}_{m',m}$
and omit the explicit angular dependence of ${}_{-2}Y_{\ell,m}$
and the precession angles in $D^{\ell}_{m',m}$.
The summation over $\ell$ and $m'$ are over the modes included in the
co-precessing frame. Here we use the {\tt PhenomHM} model~\cite{PhysRevLett.120.161102},
which contains the $(\ell, |m'|) = ((2,2), (2,1), (3,3), (3,2), (4,4), (4,3))$ modes.

Due to precession the properties of the remnant \ac{BH} in the precessing system
are different to those in the equivalent non-precessing system.
We use the same prescription as described in Sec.III.C of
Ref.~\cite{PhysRevD.100.024059}
to include the in-plane-spin contribution to the spin of the remnant \ac{BH}.
This modified final spin vector changes the ringdown spectrum of the
aligned-spin multipoles.

Lastly, we note that the models for the three ingredients
(the non-precessing model, the precession angles, and the \ac{BH} remnant model)
are independent in our construction, and can therefore
each be updated when any of them are improved.

\section{Waveform assessment}

\subsection{Mismatch Computation}
\label{sec:matches}

\begin{figure}[t!]
  \includegraphics[width=0.98\columnwidth]{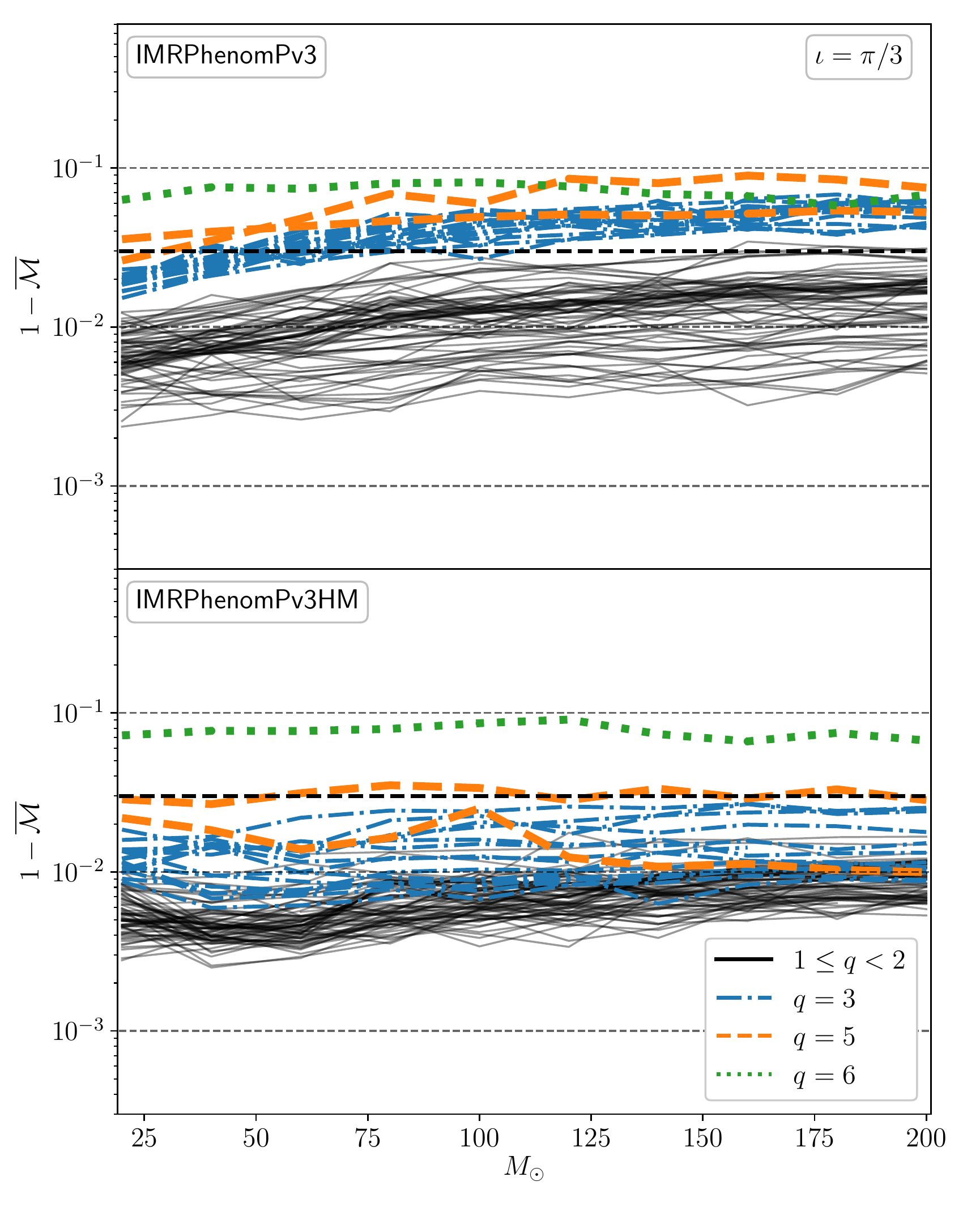}
 \caption{The results of the comparison between {\tt PhenomPv3} (top),
 {\tt PhenomPv3HM} (bottom) and the precessing \ac{NR} simulations from the public SXS catalogue.
 The figure shows the mismatch average over a reference phase and polarisation
 angle ($1 - \overline{\mathcal{M}}$) as a function of the total mass
 for an inclination of $\iota = \pi/3$.
 The worst case is SXS:BBH:0165, a short ($\sim$6 orbits) signal with mass ratio 1:6 and high precession.
 }
 \label{fig:matches}
\end{figure}

\begin{table*}[t]
\centering
\begin{tabular}{lllllll}
\hline \hline
Waveform Model                    & \multicolumn{3}{l}{{\tt PhenomPv3}} & \multicolumn{3}{l}{{\tt PhenomPv3HM}} \\ \hline
Mass-Ratio ($\#$)                          & $0$           & $\pi/3$       & $\pi/2$      & $0$     & $\pi/3$     & $\pi/2$    \\ \hline
$1 \leqslant q \leqslant 2$ (72) & $0.998^{ 0.999 }_{ 0.993 }$ & $0.989^{ 0.996 }_{ 0.977 }$ & $0.982^{ 0.993 }_{ 0.967 }$ &  $0.997^{ 0.999 }_{ 0.993 }$ & $0.993^{ 0.996 }_{ 0.986 }$ & $0.987^{ 0.992 }_{ 0.972 }$ \\
$q=3$ (15) & $0.989^{ 0.999 }_{ 0.974 }$ & $0.959^{ 0.967 }_{ 0.950 }$ & $0.941^{ 0.946 }_{ 0.933 }$ & $0.993^{ 0.997 }_{ 0.985 }$ & $0.987^{ 0.993 }_{ 0.975 }$ & $0.984^{ 0.989 }_{ 0.974 }$ \\
$q=5$ (2)  & $0.973^{ 0.978 }_{ 0.968 }$ & $0.941^{ 0.951 }_{ 0.931 }$ & $0.911^{ 0.925 }_{ 0.897 }$ & $0.990^{ 0.990 }_{ 0.989 }$ & $0.971^{ 0.975 }_{ 0.966 }$ & $0.978^{ 0.989 }_{ 0.968 }$  \\
$q=6$ (1)  & $0.863$ & $0.919$ & $0.939$ & $0.950$ & $0.914$ & $0.898$  \\ \hline \hline
\end{tabular}
\caption{Match results from Sec.~\ref{sec:matches}.
We quote the mean value of the match for each inclination
angle considered ($\iota \in [0, \pi/3, \pi/2]$ rad) and averaged
over all cases in the mass-ratio
category for the $M_{tot}=100M_\odot$ case.
The subscript and superscript are the minimum and maximum values of the
match for the mass-ratio category considered.
}
\label{table:matches}
\end{table*}

The standard metric to assess the accuracy of \ac{GW} signal models is
to calculate the noise-weighted inner product between the
\emph{template} model and an accurate \emph{signal} waveform.
As our signal we use \ac{NR} waveforms from the publicly available
SXS catalogue~\cite{PhysRevLett.111.241104,sxsonline,Boyle:2019kee} generated
using the
\ac{NR} injection infrastructure in {\tt LALSuite}~\cite{Schmidt:2017btt}.
From this catalogue we select the precessing
configurations with the highest numerical resolution. This set contains
90 systems with $q \in [1,6]$, however, the majority of cases
have $q \leqslant 3$. We have 2 cases at $q=5$ and one case at $q=6$.
There are six cases that have at least one \ac{BH} with a dimensionless
spin magnitude $|\chi| >0.5$ whereas the majority of cases have
$|\chi| \leqslant 0.5$
\footnote{During the concluding stages of this project
the SXS collaboration updated their catalgoue
to include $\sim 2000$ new simulations~\cite{Boyle:2019kee}.
We defer comparison to this catalogue to a future date.}.
For the exact list of \ac{NR}
configurations and specific details on how the mismatch calculations
were performed we refer the reader to Ref.~\cite{PhysRevD.100.024059}
where we presented an identical analysis but restricted the signals
to contain the $\ell = 2$ multipoles.

Since {\tt PhenomPv3} is constructed from {\tt PhenomD} it only
has the $\ell=|m|=2$ modes in the co-precessing frame and therefore
we expect this model to perform poorly when the contribution to the
signal due to higher modes is not negligible.
As {\tt PhenomPv3HM} is constructed from {\tt PhenomHM}
and contains the $(\ell, |m|) = ((2,2), (2,1), (3,3), (3,2), (4,4), (4,3))$
modes in the co-precessing frame we expect it to outperform {\tt PhenomPv3}.
In the \ac{NR} signal we include multipoles with $\ell \in [2,4]$ to be
consistent with the highest modeled $\ell$ mode in {\tt PhenomPv3HM}.

We use the expected noise curve for Advanced LIGO operating at design
sensitivity~\cite{aligozerodethp}
\footnote{See~\cite{aligozerodethpnew} for a more recent reference.}
with a low-frequency cutoff of $10{\rm{Hz}}$.
Due to the presence of higher modes the orbital phase of the binary is
no longer degenerate with the phase of the observed waveform, which
means the standard method to analytically maximise over the template
phase is not applicable. It is possible, however, to
analytically maximise over the template polarisation using
the skymax-SNR derived in Ref.~\cite{PhysRevD.94.024012}.
In our match calculation we analytically maximise over the template polarisation
and relative time shift and numerically optimise over the template orbital
reference phase and frequency. Finally we average the match, weighted
by the optimal SNR, over the signal orbital reference phase and polarisation
angle. See Sec.~III A in Ref.~\cite{PhysRevD.100.024059} for details.

Figure~\ref{fig:matches} shows the orientation-averaged-mismatch~\cite{PhysRevD.100.024059}
as a function of the total mass of the binary for an inclination angle $\iota = \pi/3$.
Here $\iota$ is the angle between the Newtonian orbital
angular momentum and the line of sight at the start frequency of the \ac{NR}
waveform. The first row uses the dominant multipole-only model, {\tt PhenomPv3},
and the second row uses the new higher multipole model, {\tt PhenomPv3HM},
presented here. We clearly see that for $q \geqslant 3$ that it is important
to include higher modes in the template model.

In Table~\ref{table:matches} we summarise the results of our validation study
by tabulating the results as the match (as opposed to mismatch as in
Fig.~\ref{fig:matches}) for each model according to mass ratio and inclination
angle. Next to the mass-ratio range in parentheses is the number of \ac{NR}
cases in that mass-ratio category.
Each entry in the table is calculated as follows: for the
$M_{tot} = 100 M_\odot$ case we average the match over all cases in the
mass-ratio category and write the minimum and maximum match as
subscript and superscript respectively.
\paragraph*{$1 \leqslant q \leqslant 2$:}

In this mass-ratio range both {\tt PhenomPv3} and {\tt PhenomPv3HM}
perform comparably, most likely due to the strength of
higher multipoles scaling with mass ratio.

\paragraph*{$q=3$:}

Here we start to see the importance of the higher multipoles to accurately
describe the \ac{NR} signal.
For $\iota = 0$ {\tt PhenomPv3}
has an average match of $0.989$. However, as the inclination angle increases,
thus emphasising more of the higher multipole content of the signal,
the average match drops to $0.941$ and can be as low as $0.933$.
On the other hand, {\tt PhenomPv3HM} is able to describe the \ac{NR} data
to an average accuracy of $0.984$ with a minimum value of $0.974$ for
inclined systems.

\paragraph*{$q=5$:}

At this mass-ratio the loss in performance for {\tt PhenomPv3} is
noticable even for low inclination values. At $\iota = 0$
the average match is $0.973$ dropping to $0.911$ at $\iota = \pi/2$.
The match for {\tt PhenomPv3HM} at $\iota = 0$ remains high
at $0.99$, but reduces to $0.978$ at $\iota = \pi/2$.
Note that we only have two \ac{NR} simulations at $q=5$ and are thus unable to
rigorously test the model at this and similar mass ratios.

\paragraph*{$q=6$:}

When comparing to this \ac{NR} simulation we find both models perform substantially
worse than the $q=5$ cases with even {\tt PhenomPv3} outperforming
{\tt PhenomPv3HM} with matches as low as $0.898$.
We have verified that we obtain matches of $\sim 0.97$ when restricting the
\ac{NR} waveform to just the $\ell = 2$ multipoles, consistent with
our previous study~\cite{PhysRevD.100.024059}.
We conclude that either our
model is outside its range of validity or that this \ac{NR} simulation
is inaccurate for the higher multipoles, however, our results are robust
against \ac{NR} simulations of this configuration at multiple resolutions.
This \ac{NR} simulation, SXS:BBH:0165 is exceptional for a few
reasons. First, it is a high mass-ratio system where higher multipoles
are more important. Second, it is a strongly precessing system
with primary  $\chi_1 = (0.74, 0.19, -0.5)$ and secondary
$\chi_2 = (-0.19, 0., -0.23)$ spin vectors. Finally, it is also
very short, only containing $\sim 6.5$ orbits. We encourage more
\ac{NR} simulations in this region by different \ac{NR} codes to (i) cross
check the results and (ii) populate this region with more data with which
to test and refine future models.

We conclude from our study that {\tt PhenomPv3HM} greatly improves the
accuracy towards precessing \acp{BBH} for systems with mass-ratio up to 5:1.
We expect to be able to greatly improve the accuracy and extend towards
higher mass-ratio by further calibrating the higher order multipoles
and precession effects to \ac{NR} simulations.

\subsection{Parameter Uncertainty}
\label{sec:pe}

One of the main purposes of a waveform model is to estimate the source
parameters of \ac{GW} events.
With models we can quantify the expected parameter uncertainty as a function
of the parameter space
~\cite{
  PhysRevD.95.064053,PhysRevD.95.064052,Kumar:2018hml,
  Ghosh:2015jra,
  Pankow:2016udj,
  Vitale:2014mka,Farr:2015lna,
  PhysRevD.93.084042,Stevenson:2017dlk,PhysRevD.95.064053}.
Instead of a computationally intense systematic parameter estimation campaign
we have chosen to focus on one configuration and study in detail the
dependency of parameter recovery on \ac{SNR}.
We wish to study a system where both precession and higher modes are important
and guided by previous studies~\cite{Capano:2013raa,PhysRevD.96.124024,PhysRevD.95.104038}
we chose to study a
double precessing spin, mass-ratio 3 \ac{BBH}
signal with a total mass of $150 M_\odot$ in the detector frame.
Starting at a frequency of $10$ Hz this system produces a waveform with about
20 \ac{GW} cycles and merges at a frequency of about $120$ Hz.
See Table~\ref{table:injection-pars} for specific injection values,
where $(\theta_{JN}, \phi)$ define the direction of propagation in the source
frame, $(\alpha, \delta)$ are the right-ascension and declination of the source
and $\psi$ is the polarisation angle.

We simulate this fiducial signal with {\tt PhenomPv3HM} and recover its
parameters using the parallel tempered MCMC algorithm implemented as
{\tt LALInferenceMCMC} in the publicly available LALInference
software~\cite{PhysRevD.91.042003} with {\tt PhenomPv3HM} as the template
model.
We perform three separate, zero-noise, injections to investigate how our results
depend on the injected \ac{SNR}. Specifically we inject the signal at
luminosity distances of
3000 Mpc, 1500 Mpc and 300 Mpc corresponding to a three-detector network \ac{SNR}
of 17, 35 and 176 respectively.
We use the design sensitivity noise curves for
the LIGO Hanford, LIGO Livingston and Virgo detectors~\cite{aligozerodethp}.

We present our results by tabulating the median and 90\% credible
interval on binary parameters in Tab.~\ref{table:injection-pars}
and source frame parameters in Tab.~\ref{tab:source_frame}.
We also plot the 90\% credible interval as a function of the injected \ac{SNR}
for a few chosen parameters in
Figs.~\ref{fig:90width-mass}, \ref{fig:90width-spin}
and \ref{fig:90width-spin-angle}.
In the high \ac{SNR} limit the uncertainty on the parameters
should decrease linearly wtih SNR i.e., as $1/\rho$~\cite{Vallisneri:2007ev},
which is shown as a dashed black line in these figures.

In the following discussion we
change our convention for the mass-ratio to $q \equiv m_2/m_1 \in [0, 1]$
and will abbreviate
the width of the 90\% credible interval of
parameter $X$ at an \ac{SNR} of $\rho$ as $C_{90\%}^{\rho}(X)$.

{\renewcommand{\arraystretch}{1.5}
\begin{table}
  \centering
    \begin{tabular}{p{1.7cm}p{1.7cm}p{1.7cm}p{1.7cm}l}
    \hline \hline
    Parameter                         &   \shortstack{Injection \\ Value} & \shortstack{$\rho = 17.6$ \\ $D_L = 3000$}   & \shortstack{$\rho = 35.2$ \\ $D_L = 1500$}   & \shortstack{$\rho = 176$ \\ $D_L = 300$}   \\
    \hline
    $m^{\rm{det}}_{1}/M_\odot$              & 112.500                           & $102.98^{+13.38}_{-12.26}$                   & $107.71^{+7.96}_{-7.43}$                     & $112.38^{+1.73}_{-1.75}$                   \\
    $m^{\rm{det}}_{2}/M_\odot$              & 37.500                            & $40.62^{+5.92}_{-5.29}$                      & $39.00^{+2.88}_{-2.66}$                      & $37.55^{+0.54}_{-0.52}$                    \\
    $M_{\rm{total}}^{\rm{det}} / M_\odot $  & 150.000                           & $143.64^{+11.28}_{-9.74}$                    & $146.78^{+6.48}_{-5.94}$                     & $149.93^{+1.49}_{-1.46}$                   \\
    $\mathcal{M}_{c}^{\rm{det}} / M_\odot $ & 54.940                            & $55.08^{+3.37}_{-3.20}$                      & $55.05^{+1.69}_{-1.69}$                      & $54.95^{+0.36}_{-0.34}$                    \\
    $q$                                     & 0.333                             & $0.39^{+0.11}_{-0.08}$                       & $0.36^{+0.05}_{-0.04}$                       & $0.33^{+0.01}_{-0.01}$                     \\
    $\theta_1$ / rad                              & 1.052                             & $1.14^{+0.36}_{-0.37}$                       & $1.10^{+0.27}_{-0.19}$                       & $1.05^{+0.04}_{-0.04}$                     \\
    $\theta_2$ / rad                              & 2.090                             & $1.73^{+1.01}_{-1.21}$                       & $2.04^{+0.72}_{-1.09}$                       & $2.09^{+0.14}_{-0.12}$                     \\
    $\Delta \phi_{12}$ / rad                      & 1.571                             & $2.82^{+3.11}_{-2.49}$                       & $1.79^{+3.42}_{-1.35}$                       & $1.58^{+0.24}_{-0.24}$                     \\
    $\theta_{JN}$ / rad                           & 1.050                             & $1.62^{+0.60}_{-0.73}$                       & $1.21^{+0.92}_{-0.23}$                       & $1.05^{+0.03}_{-0.03}$                     \\
    $\cos(\phi)$                            & 1.000                             & $-0.04^{+1.03}_{-0.96}$                      & $0.46^{+0.54}_{-1.46}$                       & $1.00^{+0.00}_{-0.01}$                     \\
    $\alpha$ / rad                                & 1.047                             & $4.235^{+0.124}_{-3.213}$                    & $1.070^{+3.277}_{-0.036}$                    & $1.047^{+0.004}_{-0.004}$                  \\
    $\delta$ / rad                                & 1.047                             & $-1.020^{+2.098}_{-0.125}$                   & $1.025^{+0.037}_{-2.155}$                    & $1.047^{+0.004}_{-0.004}$                  \\
    $\psi$ / rad                                  & 1.047                             & $1.52^{+0.66}_{-0.76}$                       & $1.28^{+0.80}_{-0.39}$                       & $1.05^{+0.04}_{-0.04}$                     \\
    $\chieff$                            & 0.200                             & $0.204^{+0.129}_{-0.136}$                    & $0.201^{+0.070}_{-0.074}$                    & $0.200^{+0.016}_{-0.017}$                  \\
    $\chi_{p}$                              & 0.700                             & $0.681^{+0.186}_{-0.285}$                    & $0.705^{+0.098}_{-0.105}$                    & $0.699^{+0.020}_{-0.024}$                  \\
    $|\chi_1|$                              & 0.806                             & $0.77^{+0.15}_{-0.27}$                       & $0.80^{+0.07}_{-0.09}$                       & $0.81^{+0.02}_{-0.02}$                     \\
    $|\chi_2|$                              & 0.806                             & $0.45^{+0.47}_{-0.40}$                       & $0.59^{+0.35}_{-0.42}$                       & $0.80^{+0.13}_{-0.11}$                     \\
    $D_L$ / Mpc                                   & see heading                       & $3086.50^{+739.44}_{-571.98}$                & $1465.76^{+177.84}_{-157.87}$                & $300.12^{+7.14}_{-6.94}$                   \\
     \hline \hline
    \end{tabular}
  \caption{
    Injection parameters and results from parameter estimation of
    simulated signals.
    We quote the median and 90\% credible interval.}
  \label{table:injection-pars}
\end{table}

{\renewcommand{\arraystretch}{1.5}
\begin{table*}
  \begin{tabular}{p{1.8cm}p{1.8cm}p{1.8cm}p{1.8cm}p{1.8cm}p{1.8cm}p{1.8cm}}
    \toprule
    & \multicolumn{2}{l}{\quad \quad \quad \quad $\rho$=17}          & \multicolumn{2}{l}{\quad \quad \quad \quad$\rho$=34}          & \multicolumn{2}{l}{\quad \quad \quad \quad$\rho$=176}         \\\hline
Parameter   & Inj.       & Rec. & Inj.       & Rec. & Inj.       & Rec. \\\hline
$m^{\rm{src}}_{1}/M_\odot$             &   74.56  & $67.60^{+8.22}_{-8.35}$       &   87.789 & $84.28^{+5.36}_{-5.24}$       & 105.724 & $105.54^{+1.55}_{-1.58}$  \\
$m^{\rm{src}}_{2}/M_\odot$             &   24.853 & $26.60^{+3.95}_{-3.55}$       &   29.263 & $30.48^{+2.47}_{-2.21}$       &  35.241 & $35.27^{+0.51}_{-0.50}$   \\
$M_{\rm{total}}^{\rm{src}} / M_\odot $ &   99.413 & $94.35^{+6.64}_{-7.16}$       &  117.052 & $114.82^{+3.97}_{-3.96}$      & 140.965 & $140.80^{+1.28}_{-1.27}$  \\
$M_{c}^{\rm{src}} / M_\odot $          &   36.412 & $36.11^{+2.29}_{-2.34}$       &   42.872 & $43.04^{+1.31}_{-1.28}$       &  51.631 & $51.61^{+0.31}_{-0.30}$   \\
$D_L / \rm{Mpc}$                       & 3000     & $3086.50^{+739.44}_{-571.98}$ & 1500     & $1465.76^{+177.84}_{-157.87}$ & 300     & $300.12^{+7.14}_{-6.94}$  \\
$z$                                    &    0.509 & $0.52^{+0.10}_{-0.08}$        &    0.281 & $0.28^{+0.03}_{-0.03}$        &   0.064 & $0.065^{+0.001}_{-0.001}$ \\
\hline \hline
\end{tabular}
  \caption{
    Source frame injection parameters and
    results from parameter estimation of
    simulated signals.
    We quote the median and 90\% credible interval.}
  \label{tab:source_frame}
\end{table*}

\subsubsection{Masses}

\begin{figure}[t]
  \includegraphics[width=1\columnwidth]{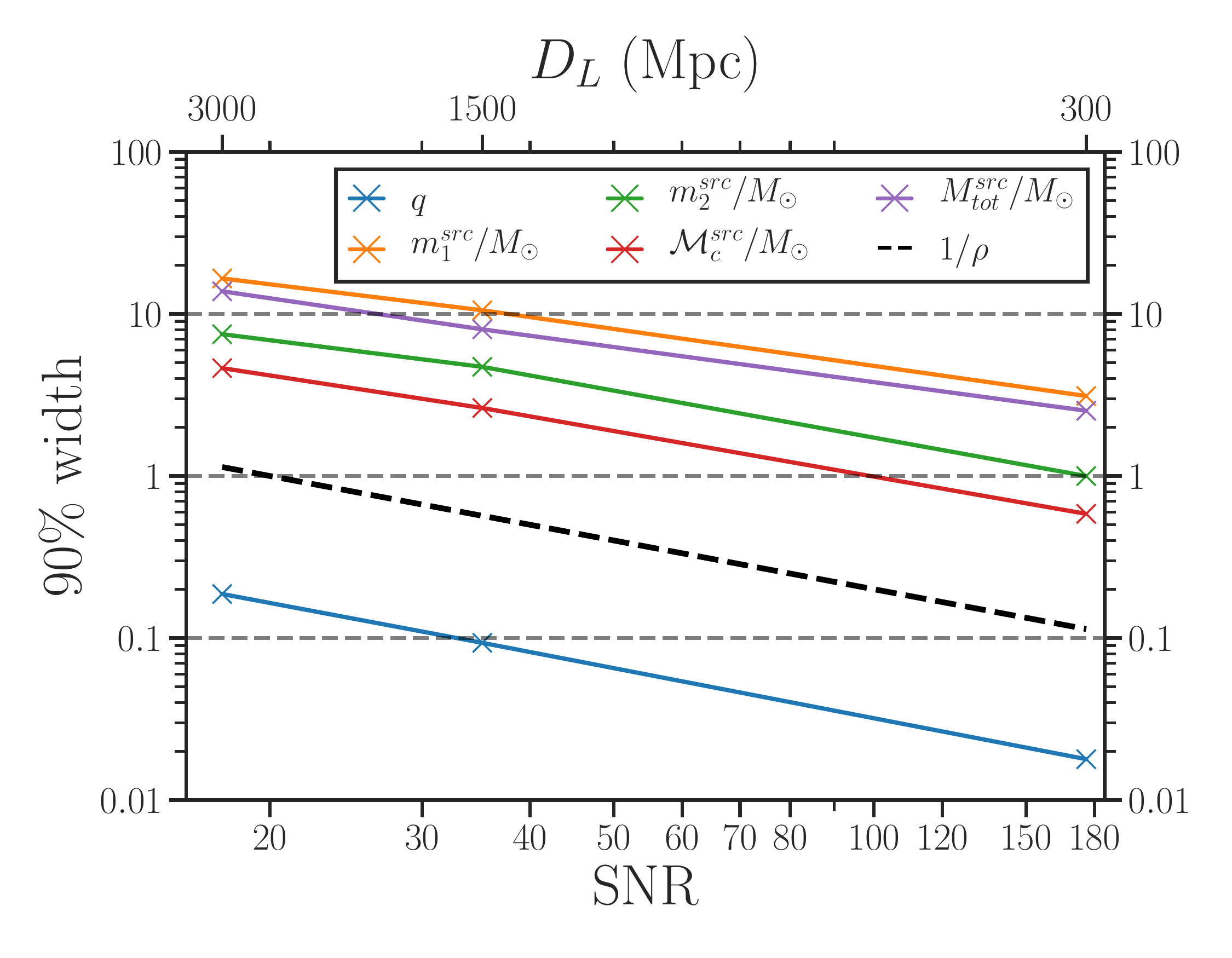}

\caption{90\% credible intervals for the source frame mass parameters as a
function of injected SNR}
\label{fig:90width-mass}
\end{figure}

Figure~\ref{fig:90width-mass} shows the source frame mass parameters;
primary mass $m^{\rm{src}}_{1}$, secondary mass $m^{\rm{src}}_{2}$,
chirp mass $\mathcal{M}_{c}^{\rm{src}}$, total mass $M_{\rm{total}}^{\rm{src}}$ and mass-ratio $q$.
We find good scaling with respect to $1/\rho$ for all source frame mass parameters.

Table~\ref{tab:source_frame} shows the injection and recovered values.
Even at the high total masses we consider here we find that the chirp mass
is still the best measured parameter with
$C_{90\%}^{17}(\mathcal{M}_{c}^{\rm{src}}) = 4.63 M_\odot$ and $C_{90\%}^{176}(\mathcal{M}_{c}^{\rm{src}}) = 0.61 M_\odot$.
The total mass is the next best measured mass parameter with low and high SNR accuracies of
$C_{90\%}^{17}(M_{\rm{total}}^{\rm{src}}) = 13.8 M_\odot$ and $C_{90\%}^{176}(M_{\rm{total}}^{\rm{src}}) = 2.55 M_\odot$ respectively.
We find the primary mass can be measured to an accuracy of
$C_{90\%}^{17}(m^{\rm{src}}_{1}) = 16.57 M_\odot$ for low SNR
and $C_{90\%}^{176}(m^{\rm{src}}_{1}) = 3.13 M_\odot$ for high SNR.
And for the secondary mass we find
$C_{90\%}^{17}(m^{\rm{src}}_{2}) = 7.5 M_\odot$
and $C_{90\%}^{176}(m^{\rm{src}}_{2}) = 1.01 M_\odot$ for low and high SNR
respectively.
Finally, we are able to constrain the mass-ratio to $C_{90\%}^{17}(q) = 0.19$
and $C_{90\%}^{176}(q) = 0.02$.

\subsubsection{Spins}

\begin{figure}[t]
  \includegraphics[width=1\columnwidth]{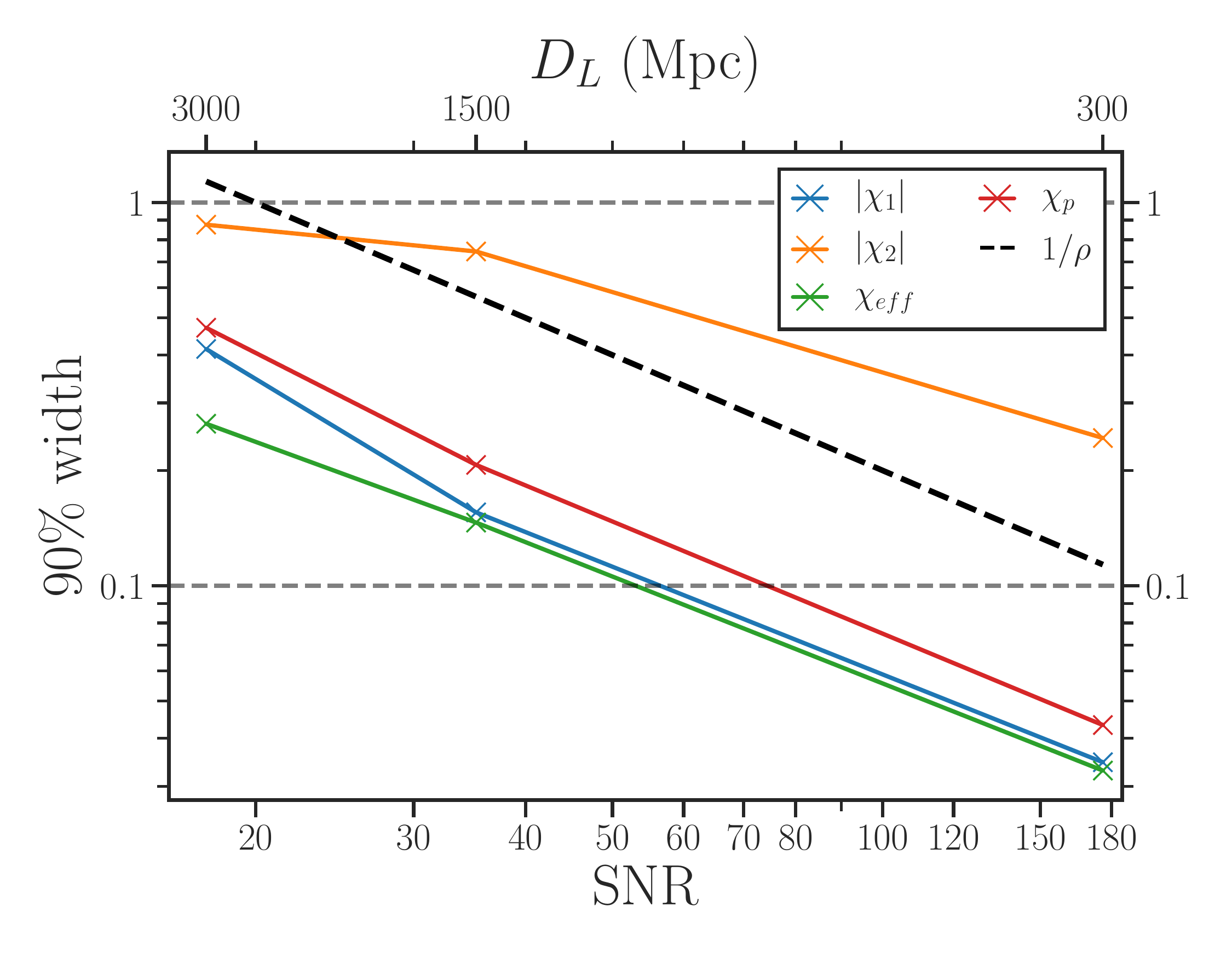}

\caption{90\% credible intervals for the \ac{BH} spin magnitudes
and effective spin parameters as a function of injected SNR.}
\label{fig:90width-spin}
\end{figure}

Figure~\ref{fig:90width-spin} shows the primary and secondary spin magnitude $|\chi_1|, |\chi_2|$,
the effective aligned-spin $\chi_{\rm{eff}}$ and effective precessing-spin $\chi_{\rm{p}}$ parameters.

With the exception of $|\chi_2|$ we find good agreement with the $1/\rho$ scaling.
This suggests that for $|\chi_2|$ the two weaker injections do not have high enough \ac{SNR}
for the posterior distribution function for this parameter to be approximated by a Gaussian~\cite{Vallisneri:2007ev}.
That being said, we do observe the 90\% width decrease with \ac{SNR} albeit at a slower rate.
At \ac{SNR} of $17$ and $34$ we find we are not able to place strong constraints on $|\chi_2|$
with $C_{90\%}^{17}(|\chi_2|) = 0.87$ and $C_{90\%}^{34}(|\chi_2|) = 0.77$.
However, at the high SNR of $176$ we begin to constrain the spin magnitude at the level of
$C_{90\%}^{176}(|\chi_2|) = 0.24$, approximately the same level of uncertainty
as $\chieff$ at a \ac{SNR} of 17. This is consistent with the study of non-precessing binaries in
Ref.~\cite{Purrer:2015nkh}, which concluded that the secondary spin will not be measurable for SNRs
below $\sim$100, but our results suggest that this carries over to precessing systems.

The primary spin magnitude is measured with much higher precision than the
secondary spin magnitude. However, constraining this parameter to a 90\% width
of less than 0.2 requires an SNR of $\sim 30$. This parameter
does follow the $1/\rho$ scaling very well and for high SNR cases we
estimate the statistical uncertainty to be $C_{90\%}^{176}(|\chi_1|) = 0.04$.

Of the effective spin parameters the effective aligned parameter $\chieff$ is the best
measured quantity.
This is closely related to the leading order spin effect in \ac{PN} theory~\cite{PhysRevD.52.848,PhysRevD.84.084037}
appearing at 1.5 PN order.
For all three SNRs the median value is always within $10^{-3}$
of the true value with the uncertainties ranging from
$C_{90\%}^{17}(\chieff) = 0.265$ to
$C_{90\%}^{176}(\chieff) = 0.033$.

Turning towards the effective precession spin parameter, $\chi_p$,
at the lowest \ac{SNR} we find the marginalised posterior
for $\chi_p$ has a median value of $0.681$, close to the true value
but with a wide uncertainty of $C_{90\%}^{17}(\chi_p) = 0.47$, spanning
almost half of the full range.
The evolution of the median value does not change significantly with increasing
\ac{SNR} however,
our measurement uncertainty does decrease with increasing \ac{SNR} as expected
and we find $C_{90\%}^{35}(\chi_p) = 0.203$ for the medium \ac{SNR} and
$C_{90\%}^{176}(\chi_p) = 0.044$ for the high \ac{SNR} case.

\begin{figure}[t]
  \includegraphics[width=1\columnwidth]{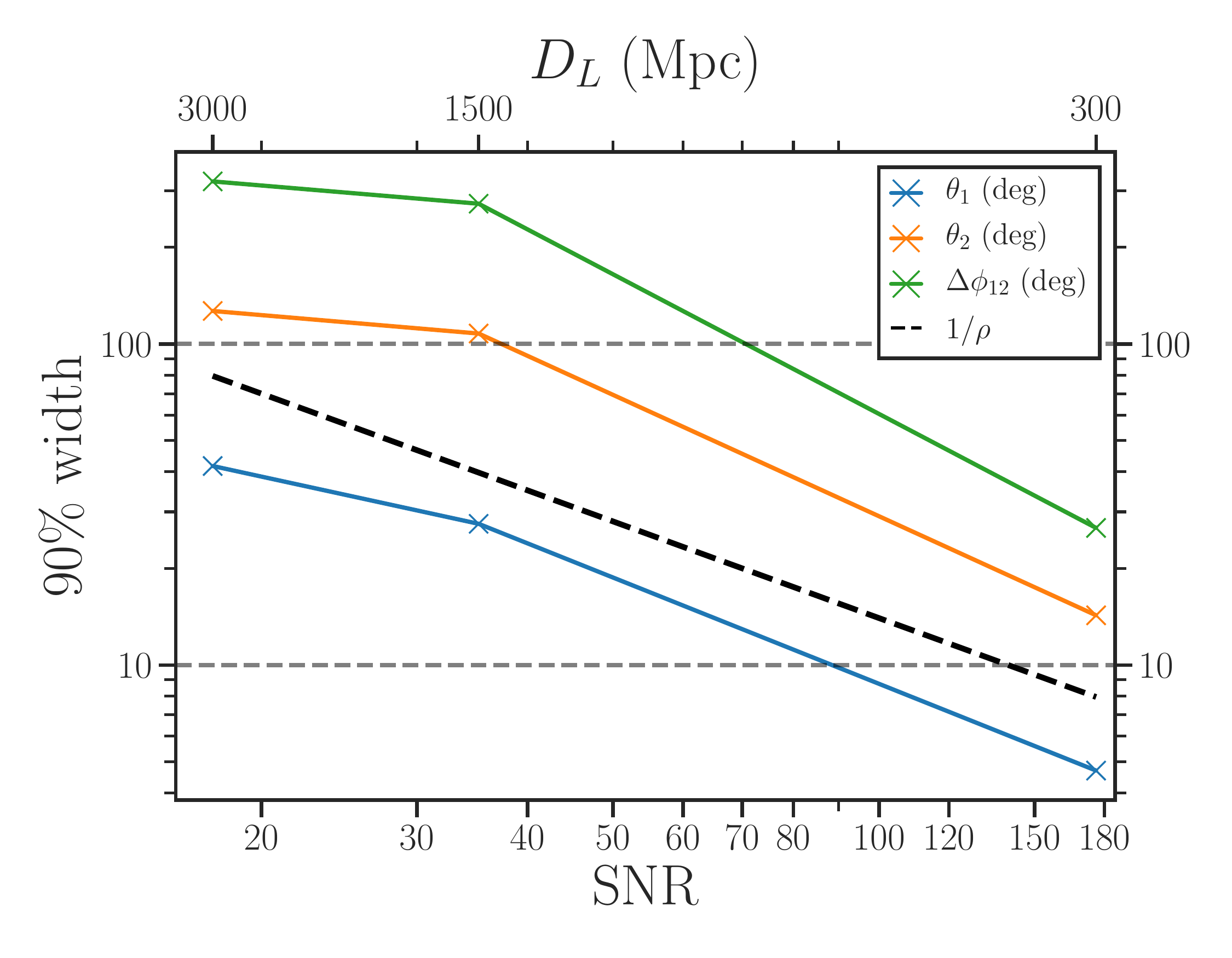}
\caption{90\% credible intervals for the \ac{BH} spin orientation
parameters as a function of injected SNR.}
\label{fig:90width-spin-angle}
\end{figure}

Figure~\ref{fig:90width-spin-angle} shows the spin orientation parameters.
$\theta_1$ and $\theta_2$ are the polar angles of the primary and secondary spin vectors
with respect to the orbital angular momentum at the reference frequency.
The angle $\Delta \phi_{12}$ is the angle between the primary and secondary spin vectors
projected into the instantaneous orbital plane at the reference frequency.
This angle is particularly useful when characterising precessing binaries
as $\Delta \phi_{12} = 0 $ or $\Delta \phi_{12} = \pi $ are resonant spin configurations
(if other conditions on the mass-ratio and spin magnitudes are met)~\cite{Schnittman:2004vq}.

We find $\theta_1$ has good \ac{SNR} scaling with
$C_{90\%}^{17}(\theta_1) = 0.73$ rad ($\sim 42$ deg)
and $C_{90\%}^{176}(\theta_1) = 0.08$ rad ($\sim 5$ deg).
Furthermore, $\theta_2$ and $\Delta \phi_{12}$ are measured much less accurately and
require SNRs of $\sim 60$ and $\sim 100$ to achieve statistical uncertainties
of $\sim 1$ rad ($\sim 60$ deg), respectively.
However, in the event of a high SNR signal we find we are able to constrain
$\theta_2$ to $C_{90\%}^{176}(\theta_2) = 0.26$ rad ($\sim 15$ deg)
and $\Delta \phi_{12}$ to $C_{90\%}^{176}(\Delta \phi_{12}) = 0.48$ rad ($\sim 28$ deg).

In summary, we find that the primary spin magnitude $|\chi_1|$ and polar angle
$\theta_1$ can be constrained at an \ac{SNR} of $\sim 30$,
while the seconday spin magnitude $|\chi_2|$ and polar angle
$\theta_2$, as well as the the information about the relative orientation of
the spin vectors $\Delta \phi_{12}$ are not constrained until we reach an
\ac{SNR} of $\sim 200$.

\subsubsection{Waveform Systematics}
\label{sec:wfsys}

Parameter estimation on a \ac{GW} event with a waveform model
that does not include relevant physics effects could result in biased results.
To quantify the size of the bias due to neglecting
higher modes and/or precession for this signal we repeat our
parameter estimation analsis with four additional models.

The waveform models we use are listed in Tab.~\ref{tab:models},
where we mark whether or not each model contains precession and/or higher modes.
{\tt PhenomD} is the baseline model upon which the other {\tt Phenom} models
used in this work are built. We include two different precessing models
{\tt PhenomPv2} and {\tt PhenomPv3} to gauge systematics on precession.
{\tt PhenomHM} includes higher modes but is a non-precessing model and finally
the precessing and higher mode model {\tt PhenomPv3HM} presented in this article.

Our results are presented in Fig.~\ref{fig:injres}. From left to
right the columns show the one dimensional marginalised posterior
distribution for the; $m_1^{det}$, $m_2^{det}$, $\chi_{\rm{eff}}$
and $\chi_{\rm{p}}$.
The rows from top to bottom show the results for the low $(\rho=17)$,
medium $(\rho=34)$ and high $(\rho=176)$ SNR injections.
The true value is shown as a vertical dashed black line.
For all SNRs we find biases in the recovered masses for all models other than
PhenomPv3HM i.e., the model that was used to produce the synthetic signal.
This suggests that for real GW signals that are similar to this injection
require analysis with models that contain \emph{both} the effects
of precession and higher modes.
For the high SNR case multi-mode posteriors are found for the {\tt PhenomD} case.
For $\chi_{\rm{eff}}$ we find that for the low SNR injection
the true value is within the $90 \%$ credible interval (CI) and therefore not considered biased
however, as the SNR of the injection is increased we find that $\chi_{\rm{eff}}$
can become heavily biased for the two precessing models
but remains unbiased for the non-precessing models.
For $\chi_{\rm{p}}$ we find that the precessing and non-higher mode models
({\tt PhenomPv2} and {\tt PhenomPv3}) consistently favour larger values of $\chi_{\rm{p}}$
as the SNR increases.
Interestingly we also start to find large differences between
{\tt PhenomPv2} and {\tt PhenomPv3} at the high SNR cases.

\begin{table}
\begin{tabular}{l l l}
\toprule
\textbf{Model} & Precession & Higher Modes \\
{\tt PhenomD} \cite{PhysRevD.93.044006,Khan:2015jqa} &\xmark &\xmark \\
{\tt PhenomPv2} \cite{Hannam:2013oca,PhysRevD.91.024043} &\cmark &\xmark \\
{\tt PhenomPv3} \cite{PhysRevD.100.024059} &\cmark &\xmark \\
{\tt PhenomHM} \cite{PhysRevLett.120.161102} &\xmark &\cmark \\
{\tt PhenomPv3HM}  & \cmark & \cmark  \\ \hline \hline
\end{tabular}
\caption{Waveform models that we use to analyse GW170729 and highlighting
which physical effects are included each model.}
\label{tab:models}
\end{table}

\begin{figure*}
\includegraphics[width=0.5\columnwidth]{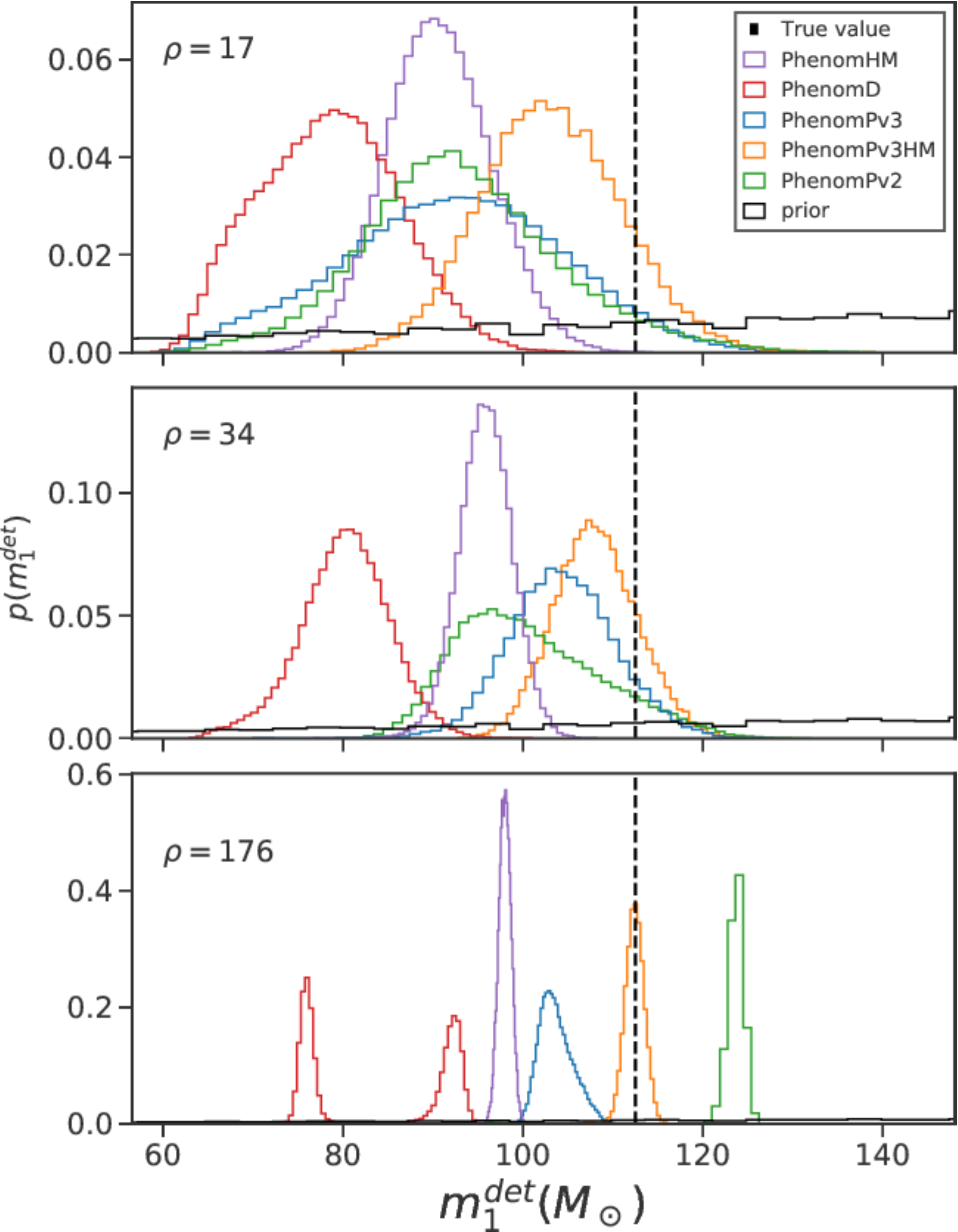}
\includegraphics[width=0.5\columnwidth]{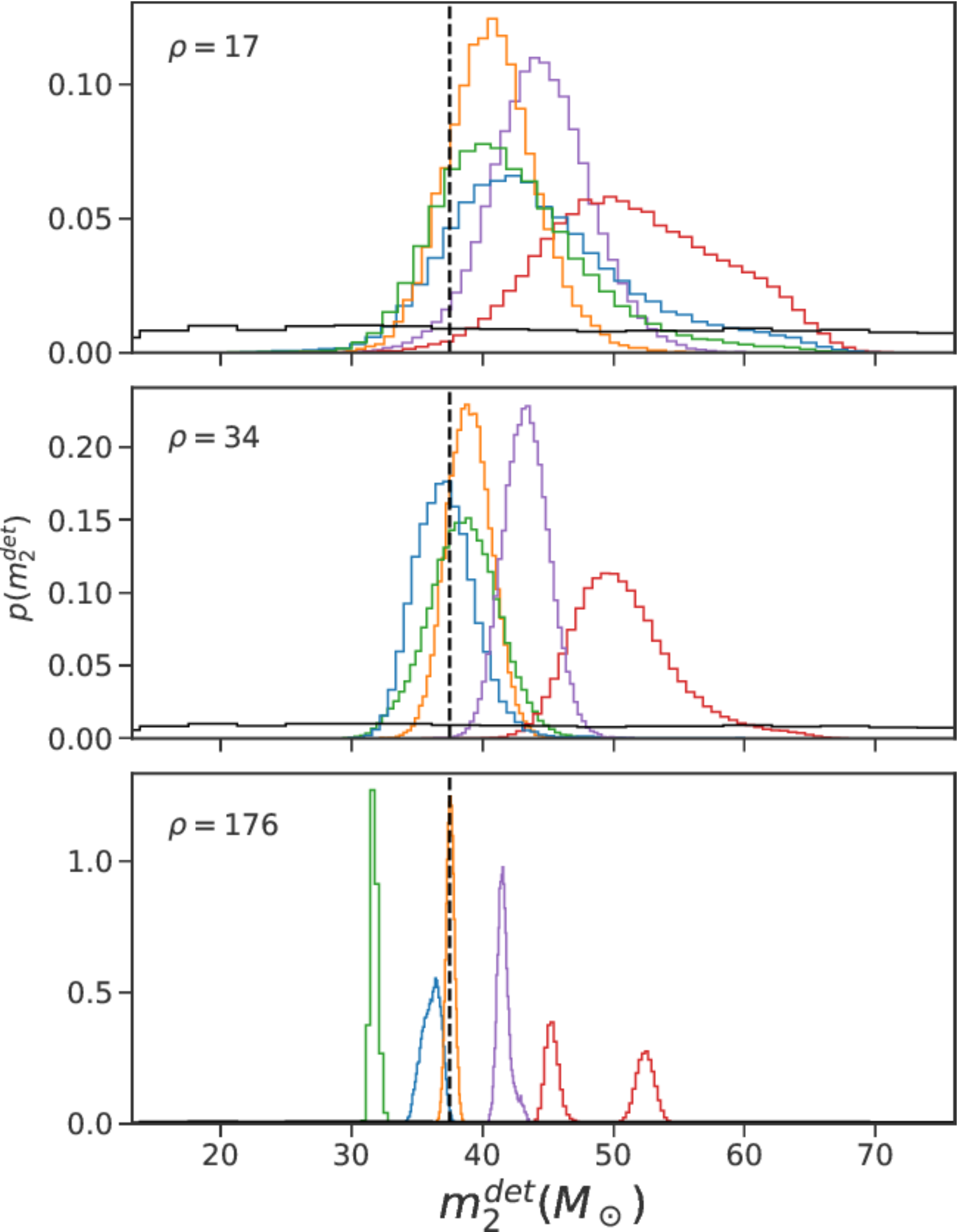}
\includegraphics[width=0.5\columnwidth]{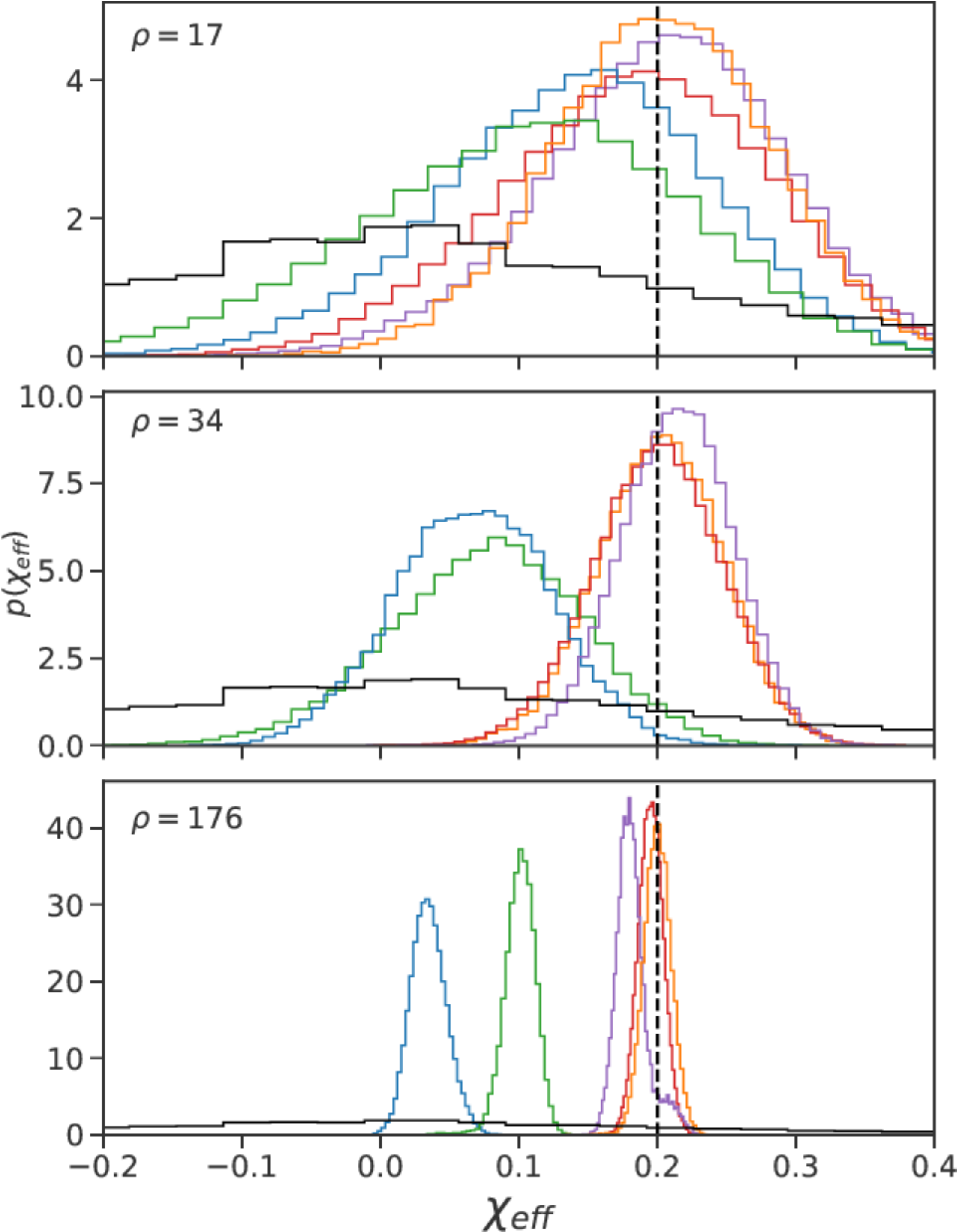}
\includegraphics[width=0.48\columnwidth]{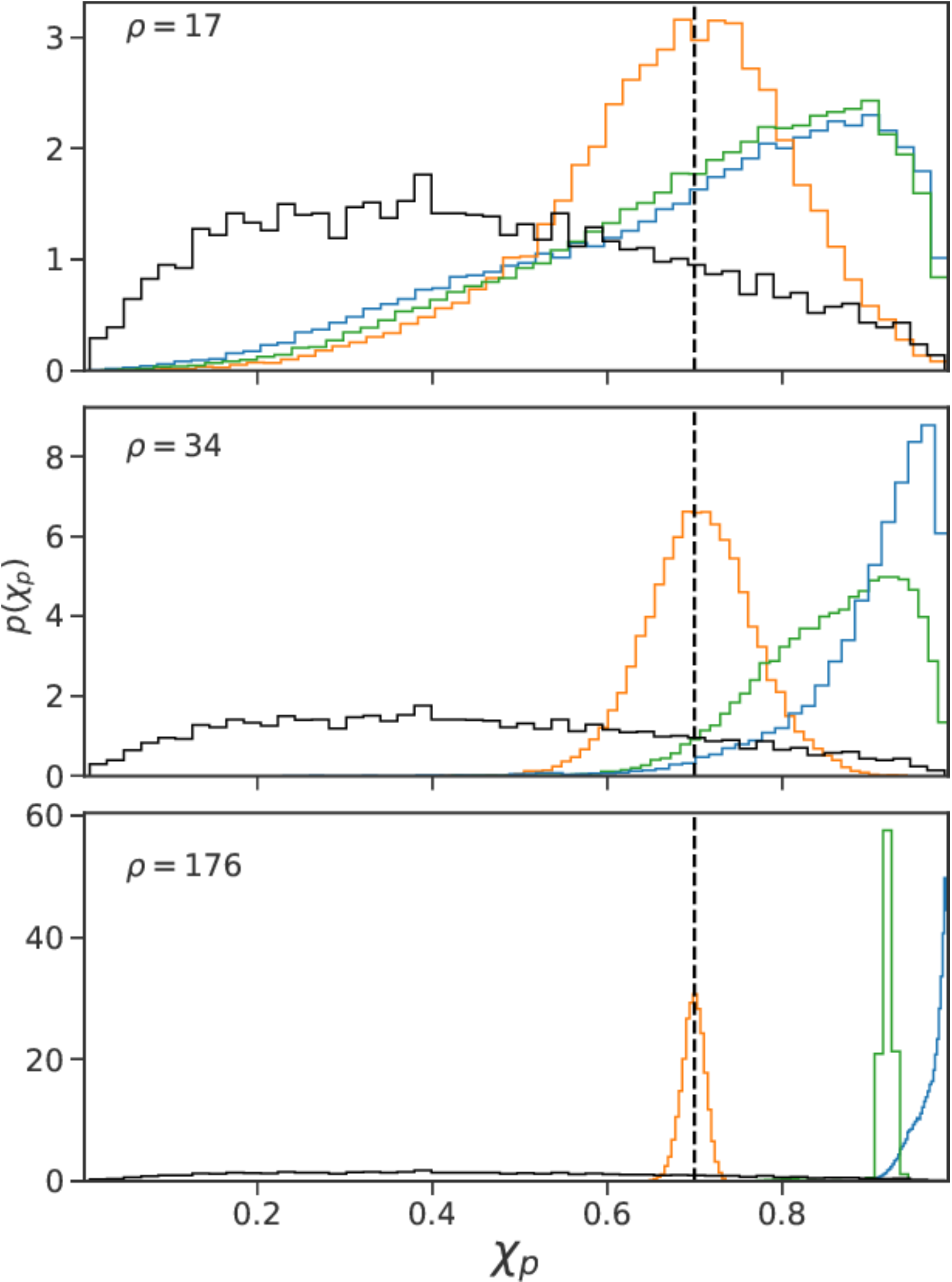}
\caption{One dimensional marginal posterior probability distributions
for detector-frame primary and secondary masses (1st and 2nd columns respectively),
effective aligned-spin $\chi_{\rm{eff}}$ and effective precession spin $\chi_{\rm{p}}$
parameters (3rd and 4th columns respectively).
Each row, from top to bottom, shows results for the low $(\rho=17)$,
medium $(\rho=34)$ and high $(\rho=176)$ SNR injections.
The true value is marked as a vertical black dashed line.
The prior is shown as a black historgram.
We show results for {\tt PhenomD} (red), {\tt PhenomHM} (purple), {\tt PhenomPv2} (green),
{\tt PhenomPv3} (blue) and {\tt PhenomPv3HM} (orange).
Note the results for $\chi_{\rm{p}}$ do not show {\tt PhenomD} or {\tt PhenomHM}
as they are aligned-spin models only.
}
\label{fig:injres}
\end{figure*}

\subsection{GW170729 Analysis}
\label{sec:gw170729}

Of the ten binary-black-hole observations reported by the LIGO-Virgo collaborations~\cite{LIGOScientific:2018mvr},
GW170729 shows the strongest evidence for unequal masses, making it the most
likely signal for which higher modes could impact parameter measurements.
This motivated the study in Ref.~\cite{Chatziioannou:2019dsz}, where the authors analysed GW170729 with
two new aligned-spin and higher mode models ({\tt SEOBNRv4HM}~\cite{PhysRevD.98.084028}
and {\tt PhenomHM}~\cite{PhysRevLett.120.161102}). They found that the
models preferred to interpret the data as the \ac{GW} signal coming from
a higher mass-ratio system with estimates for the mass-ratio changing from
$0.62^{+0.36}_{-0.23}$ for {\tt PhenomPv2} to $0.52^{+0.26}_{-0.21}$ for {\tt PhenomHM}
($90 \%$ credible interval). This event also has evidence for a
positive $\chieff$, although when analysed with higher modes
the $90 \%$ credible interval for $\chieff$ extended to include zero.
This event has also been analysed in~\cite{2019arXiv190505477P}
with the aligned-spin and higher mode model {\tt NRHybSur3dq8}~\cite{Varma:2018mmi} where the the authors draw similar conclusions.
Motivated by this we prioritise GW170729 to
analyse first with {\tt PhenomPv3HM} and compare to existing results.
We use the posterior samples for {\tt PhenomHM} from~\cite{Chatziioannou:2019dsz},
and for {\tt PhenomPv2} from~\cite{gwtc1samples}.
Results for {\tt PhenomD}, {\tt PhenomPv3} and {\tt PhenomPv3HM}
were computed for this work using the {\tt LALInferenceMCMC} code~\cite{Veitch:2014wba}.

In Fig.~\ref{fig:gw170729results} we show the joint posterior for
the source frame component masses ($m^{\rm{src}}_{1}$, $m^{\rm{src}}_{2}$)
in the upper left;
the aligned effective spin and mass-ratio ($\chieff$, $q$) in the upper right
and finally the luminosity distance and inclination angle ($D_L$, $\iota$)
in the bottom plot.
The quantitative parameter estimates for the source properties are provided
in Tab.~\ref{table:gw170729table}.
Our posterior on the effective precession parameter $\chi_p$ is consistent
with previous results and shows no significant differences due to
different choice of precession model (between {\tt PhenomPv2} and {\tt PhenomPv3})
or including both precession and higher modes as in {\tt PhenomPv3HM}.
We find that the marginal posterior effective aligned-spin parameter $\chieff$
and luminosity distance $D_L$ are
remarkably similar to the results from {\tt PhenomHM}.
The posterior for the inclination angle $\iota$ for {\tt IMPhenomPv3HM}
has more support for more inclined viewing angles; however, the
change is minor.

Interestingly, not only do we find remarkably consistent results between
{\tt PhenomD} and {\tt PhenomPv2} as discussed in~\cite{Chatziioannou:2019dsz}
but also with {\tt PhenomPv3}. This indicates that precession alone
does not influence our inference for this event.
However, including precession in addition to higher modes in the analysis
does noticeably shift the posterior, albeit not very significantly
in terms of the 90\% CIs, which mostly overlap.

We find that the one-dimensional marginal posterior for the mass-ratio is
pushed further
towards lower mass-ratio values (more asymmetric) when using {\tt PhenomPv3HM}
where we find $q=0.47^{+0.28}_{-0.16}$ ($90\%$ level),
implying that including precession \emph{and} higher modes reinforces the
findings of ~\cite{Chatziioannou:2019dsz}.
As more asymmetric masses are favoured the estimate for
the primary mass (source frame) is shifted towards higher values
and the secondary mass is shifted towards lower values
where we find $m_1^{src} = 58.25^{+11.73}_{-12.53} M_\odot$
and $m_2^{src} = 28.18^{+9.83}_{-7.65} M_\odot$.

By favouring larger mass estimates for the primary \ac{BH} we challenge
formation models to describe this event through
standard stellar evolution mechanisms.
In particular our results inform the
pulsational pair-instability supernova (PPISN)
mechanism~\cite{10.1093/mnras/stv3002, Woosley_2017}.
The population synthesis analysis in~\cite{2019arXiv190402821S}
investigated the resulting distribution of \ac{BH} masses subject to
different PPISN models.
They find that in three out of the four models that they explore, the
maximum \ac{BH} mass is $\sim 40 M_\odot$
~\cite{refId0,2018arXiv181013412M,2019ApJ...878...49W},
and in one of the models the maximum \ac{BH} mass is $\sim 58 M_\odot$
~\cite{Woosley_2017}.
In Fig.~\ref{fig:gw170729m1} we show the one-dimensional marginal posterior
for the source-frame primary mass resulting from the analysis using
{\tt PhenomPv2} (blue), {\tt PhenomHM} (orange) and {\tt PhenomPv3HM} (purple).
The 90\% credible interval of each result is shown as the shaded area under
their respective curves.
The vertical black dashed lines denote the maximum
\ac{BH} mass from the four different PPISN models that were investigated
in~\cite{2019arXiv190402821S}.
We do not show the posterior for {\tt PhenomD} or {\tt PhenomPv3}
as these are consistent with the {\tt PhenomPv2} posterior.

When using {\tt PhenomPv2} to analyse the data we find that the
maximum \ac{BH} mass for all PPISN models are consistent with the
posterior. When we include non-precessing, higher modes ({\tt PhenomHM})
the PPISN models that predict maximum \ac{BH} masses of $\sim 40 M_\odot$
~\cite{refId0,2018arXiv181013412M,2019ApJ...878...49W} are excluded at the
following level.
In the posterior $1.3 \%$ of samples have a mass of $\leqslant 40 M_\odot$.
As noted previously when we include both precession and higher modes
the primary mass shifts slightly higher resulting in
$0.6 \%$ of samples having a mass of $\leqslant 40 M_\odot$.
If we assume that the primary \ac{BH} in the GW170729 binary underwent
a PPISN then the following PPISN
models~\cite{refId0,2018arXiv181013412M,2019ApJ...878...49W}
are disfavoured at greater than 90\% credibility
and the maximum \ac{BH} mass as predicted by~\cite{Woosley_2017}
is consistent with our results.
There are some caveats to these results however.
In~\cite{2019arXiv190402821S} the authors uses a linear fit to the PPISN
model of~\cite{Woosley_2017} that systematically predicts
larger remnant \ac{BH} masses for pre-supernova helium (core) masses
$M_{\rm{He}} > 60 M_\odot$ than the model of~\cite{Woosley_2017} predicts.
This in turn leads to larger maximum \ac{BH} masses for this particular model.
However, the size of this systematic uncertainty is unknown.
Another caveat in the analysis of~\cite{2019arXiv190402821S}
is that the models \cite{2018arXiv181013412M, 2019ApJ...878...49W} have
an uncertianty of $\sim 5 M_\odot$ to account for the difference
between the gravitational and baryonic mass~\cite{2012ApJ...749...91F}.

\begin{figure*}
\includegraphics[width=0.88\columnwidth]{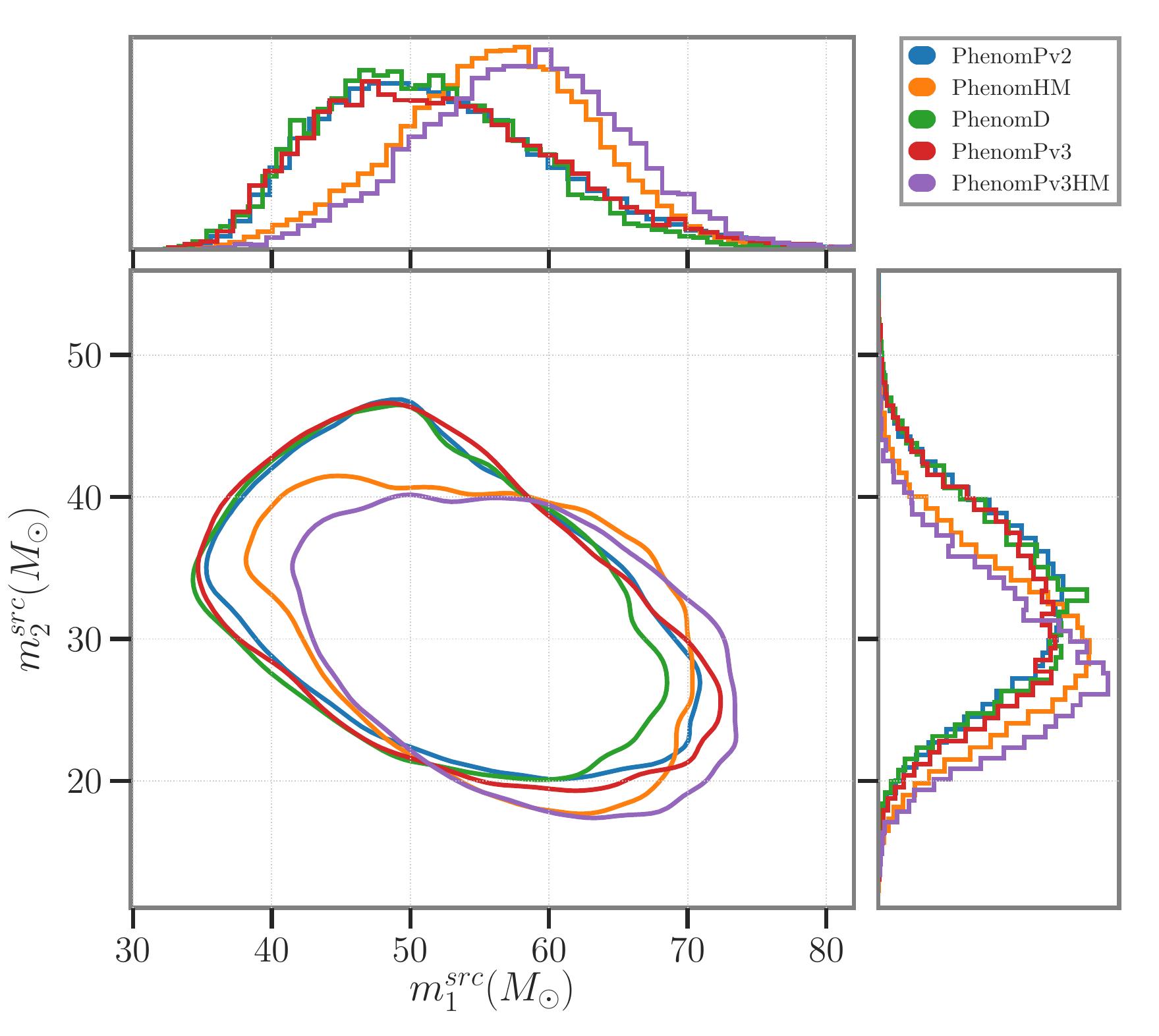}
\includegraphics[width=0.92\columnwidth]{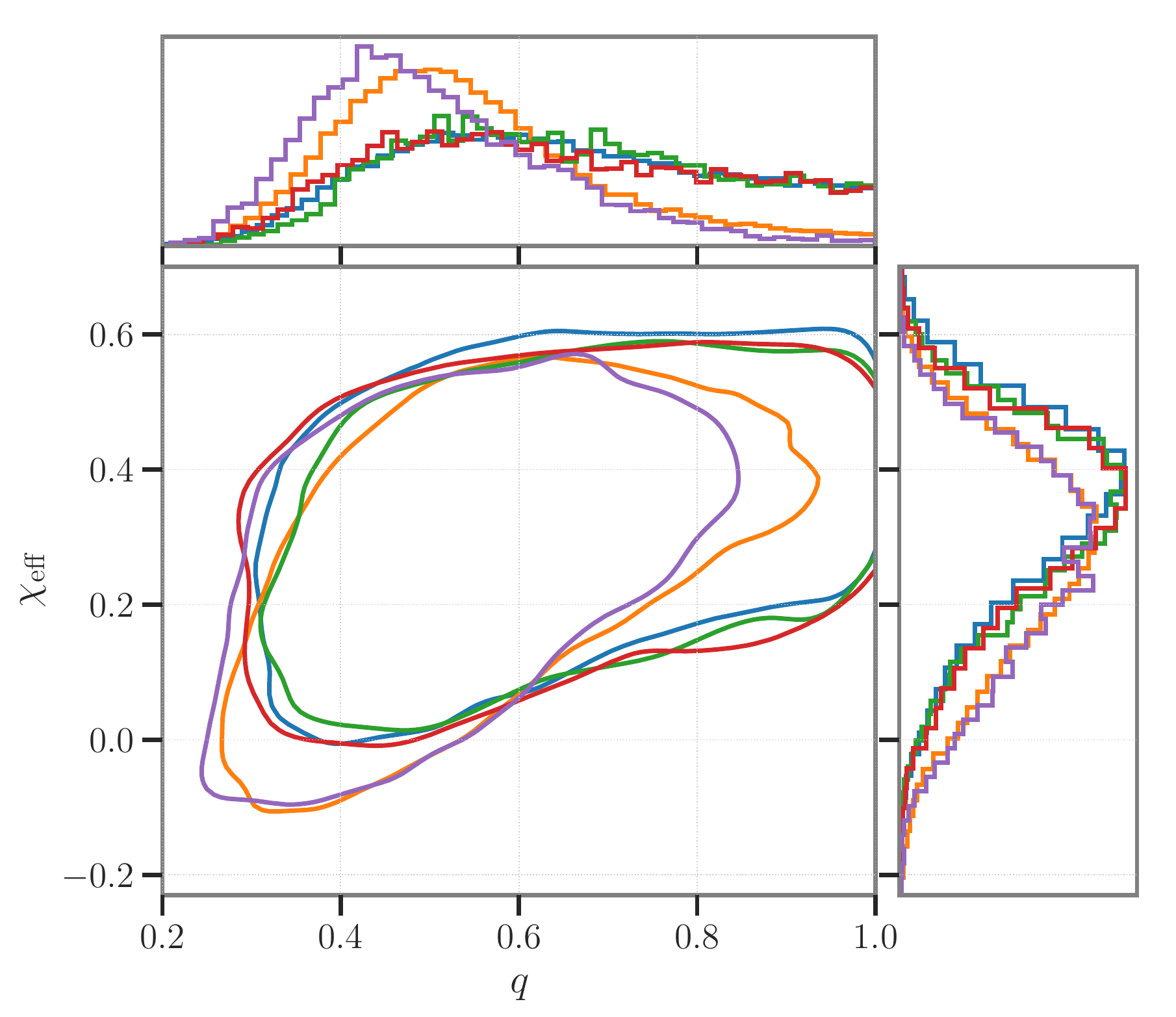}
\includegraphics[width=0.92\columnwidth]{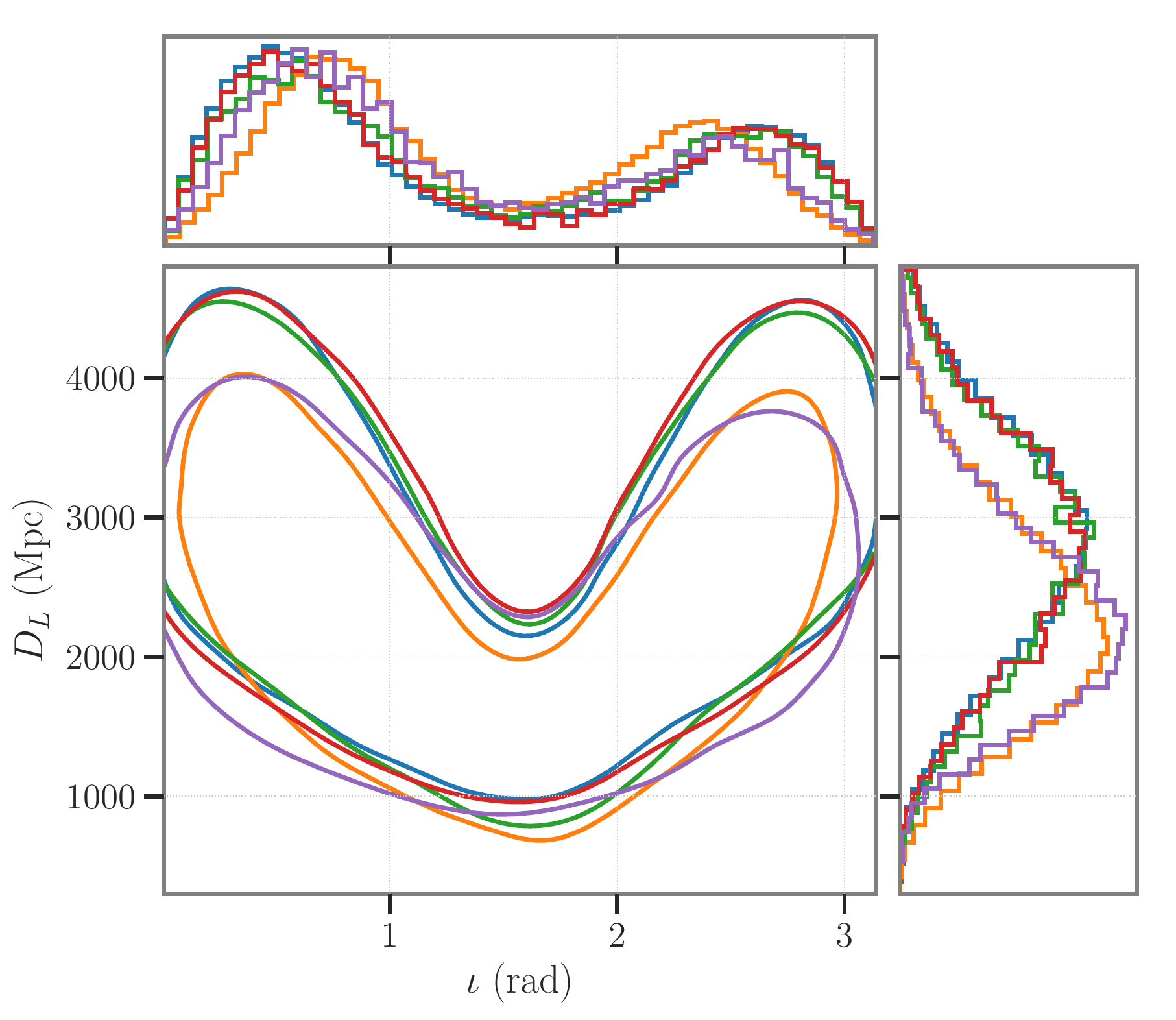}
\caption{GW170729 parameter estimation results.
Top left: component source-frame masses ($m^{\rm{src}}_{1}$, $m^{\rm{src}}_{2}$).
Top right: effective spin and mass-ratio ($\chieff$, $q$).
Bottom: inclination angle and luminosity distance ($\iota$, $D_L$).
The contour lines correspond to a credible level of $90\%$.
}
\label{fig:gw170729results}
\end{figure*}

\begin{table*}
\centering
\begin{tabular}{llllll}
\hline
Parameter                                                 & {\tt PhenomD}                         & {\tt PhenomHM }                      & {\tt PhenomPv2  }                     & {\tt PhenomPv3 }                      & {\tt PhenomPv3HM }                   \\
\hline
Primary Source Mass: $m^{\rm{src}}_{1}/M_\odot$           & $50.55^{+14.02}_{-10.64}$ & $56.36^{+11.08}_{-12.41}$ & $51.22^{+16.19}_{-10.99}$ & $51.39^{+16.35}_{-11.58}$ & $58.25^{+11.73}_{-12.53}$ \\
Secondary Source Mass: $m^{\rm{src}}_{2}/M_\odot$         & $32.18^{+10.18}_{-8.84}$  & $29.45^{+9.72}_{-8.36}$   & $32.43^{+9.75}_{-9.46}$   & $31.67^{+10.43}_{-9.27}$  & $28.18^{+9.83}_{-7.65}$   \\
Total Source Mass: $M_{\rm{total}}^{\rm{src}} / M_\odot $ & $82.80^{+15.29}_{-10.82}$ & $85.16^{+14.00}_{-10.53}$ & $83.93^{+14.74}_{-10.91}$ & $83.52^{+14.94}_{-11.09}$ & $86.18^{+13.42}_{-10.77}$ \\
Mass-Ratio: $q$                                           & $0.64^{+0.31}_{-0.24}$    & $0.52^{+0.31}_{-0.18}$    & $0.63^{+0.32}_{-0.26}$    & $0.62^{+0.33}_{-0.26}$    & $0.48^{+0.28}_{-0.16}$    \\
Effective Aligned Spin: $\chi_{\rm{eff}}$                 & $0.34^{+0.19}_{-0.26}$    & $0.28^{+0.22}_{-0.28}$    & $0.36^{+0.19}_{-0.28}$    & $0.34^{+0.19}_{-0.27}$    & $0.27^{+0.21}_{-0.28}$    \\
Effective Precession Spin: $\chi_{\rm{p}}$                & N/A                       & N/A                       & $0.44^{+0.35}_{-0.29}$    & $0.44^{+0.36}_{-0.30}$    & $0.42^{+0.39}_{-0.29}$    \\
Luminosity Distance: $D_L$ / Mpc                          & $2749^{+1353}_{-1359}$    & $2241^{+1391}_{-1065}$    & $2831^{+1371}_{-1340}$    & $2797^{+1386}_{-1318}$    & $2270^{+1307}_{-974}$     \\
redshift: $z$                                             & $0.48^{+0.19}_{-0.21}$    & $0.40^{+0.20}_{-0.17}$    & $0.49^{+0.19}_{-0.21}$    & $0.48^{+0.19}_{-0.20}$    & $0.41^{+0.19}_{-0.16}$    \\
\hline
\end{tabular}
\caption{Parameter estimation results for GW170729. Masses are quoted in the source frame.
We quote the median and the 90\% symmetric credible interval of the
one-dimensional marginalised posterior distributions.
}
\label{table:gw170729table}
\end{table*}

\begin{figure}
\includegraphics[width=\columnwidth]{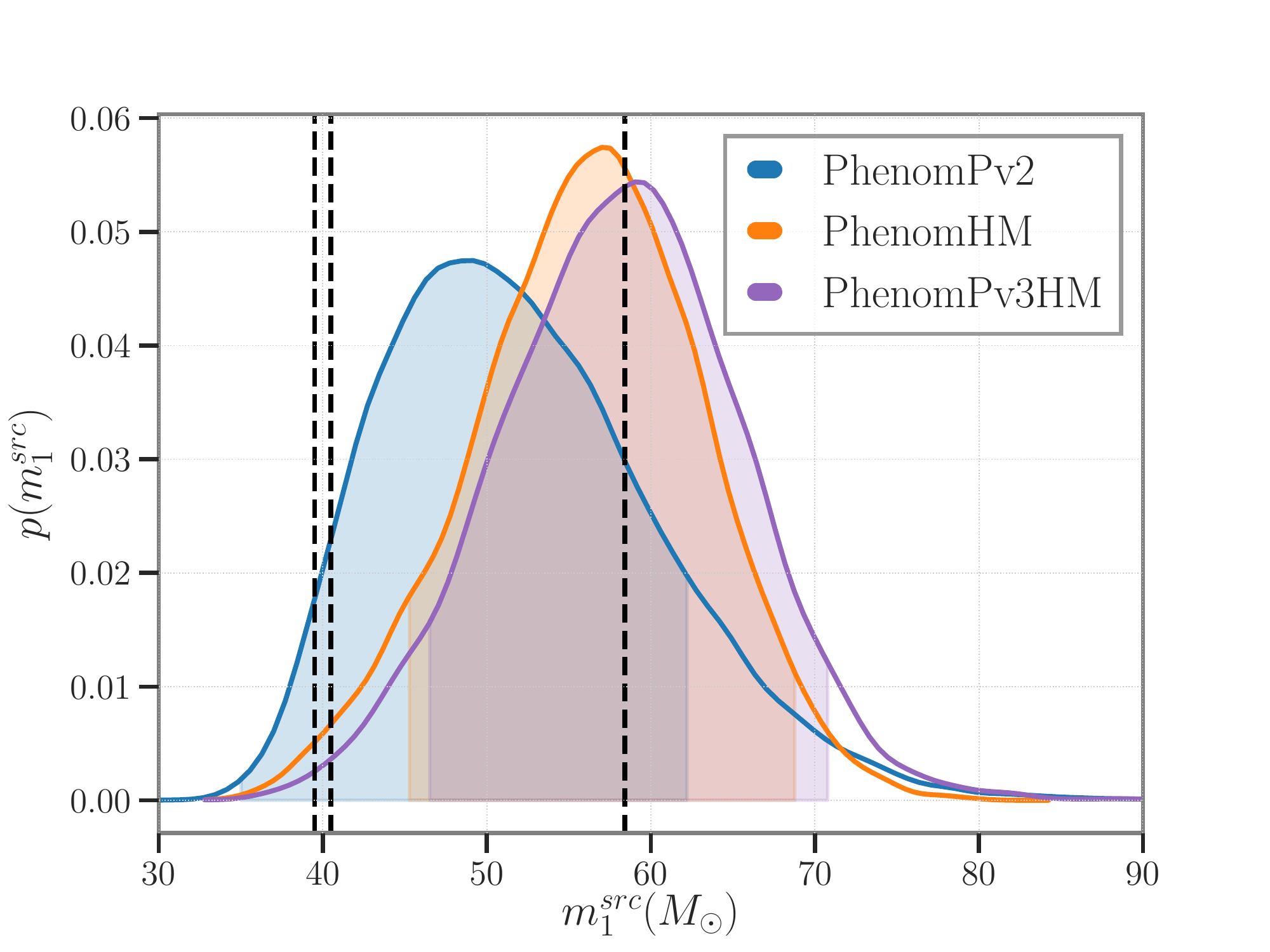}
\caption{one-dimensional marginal posterior distribution for the primary
source-frame mass.
The posteriors for three waveform models are shown: {\tt PhenomPv2} (blue),
{\tt PhenomHM} (orange) and {\tt PhenomPv3HM} (purple).
The 90\% credible interval of each result is shown as the shaded area under
their respective curves.
We also plot as vertical black dashed lines the maximum
\ac{BH} mass from four different PPISN models, which were investigated
in Ref.~\cite{2019arXiv190402821S}.
Ref.~\cite{Woosley_2017} predict a maximum mass of $58.4 M_\odot$,
Ref.~\cite{refId0} predicts $40.5 M_\odot$ and
Refs.~\cite{2018arXiv181013412M,2019ApJ...878...49W} both predict
$39.5 M_\odot$.
}
\label{fig:gw170729m1}
\end{figure}

\section{Discussion and Future}

In this work we have presented
the first, frequency-domain, phenomenological \ac{IMR} model for
spin-precessing \acp{BBH} that also includes the effects of subdominant
multipoles --- beyond the quadrupole --- in the co-precessing frame.
By comparing to a large set of precessing \ac{NR} simulations we find that
our simple model is able to accurately reproduce the expected \ac{GW} signal
with an accuracy of $99\%$ ($97\%$) for small (high) inclinations, a significant
improvement over models that do not include subdominant multipoles, which
have accuracies of $97\%$ ($91\%$) for small (high) inclinations.

Precise measurements of \ac{BH} spins from \ac{GW} observations
requires high \ac{SNR} events in part due to
the relatively high \ac{PN} order
that spin effects appear at.
We performed an idealized parameter estimation analysis
to quantify the precision to which the \ac{BH} spin magnitude and orientation
can be measured, ignoring any effects of systematic error
on the waveform.
We find, for this particular system (See Table~\ref{table:injection-pars}),
that the primary spin
parameters are more tightly constrained than the secondary spin,
as expected for an unequal-mass system such as this.
In the following discussion we remind the reader that the low, medium and
high SNR cases have corresponding values of $17$, $34$ and $176$ respectively.
The primary spin magnitude can be constrained to a 90\% CI of
$0.42$ for the low SNR case (about half the width of the physical range)
and to a 90\% CI of $0.04$ for the high SNR case.
The secondary spin magnitude cannot be meaningfully constrained until the
high SNR case with a 90\% CI of $0.24$.
The primary spin polar angle shows reasonably good agreement with the
expected SNR scaling and can be constrained to $\sim 42$ deg (low SNR)
and $\sim 5$ deg (high SNR) at 90\% CI.
The secondary spin polar angle shows poor agreement with the
expected SNR scaling and we find can only be meaningfully constrained
($\sim 15$ deg) for the high SNR case.
The azimuthal angle between the spins ($\Delta \phi_{12}$)
shows poor scaling with \ac{SNR}. We find that only the highest \ac{SNR}
case was able to constrain $\Delta \phi_{12} \lesssim 28$ deg.
Our parameter estimation study is only a point estimate for the
size of the uncertainty on binary properties and
a systematic study that explores the parameter space of precessing binaries
is required to draw more general
conclusions~\cite{PhysRevD.95.064053,PhysRevD.95.064052}.
However, recent work in understanding precession better
may help make such a study tractable by focusing on regions where
we expect precession to
be measurable~\cite{2019arXiv190800555F, 2019arXiv190805707F}.

We have analysed the \ac{GW} event GW170729 with the new
precessing and higher mode model.
We have shown that while the general interpretation of this event
is unchanged we find that even small shifts in the posteriors due
to using different waveform models, with different physical effects
incorporated, can be enough to inform astrophysical
models such as the PPISN mechanism as we considered in this paper.
If we assume that the primary \ac{BH} in the GW170729 binary underwent
a PPISN then we disfavour the PPISN
models from \cite{refId0,2018arXiv181013412M,2019ApJ...878...49W}
at greater than 90\% credibility and our results are consistent with~\cite{Woosley_2017}.
See~\cite{Farmer:2019jed} for a recent investigation into the location
of the PPISN model mass-gap.

Our model is analytic and natively in the frequency-domain, and as such it can
be readily used in likelihood acceleration methods such as \ac{ROQ}~\cite{PhysRevD.94.044031}
or ``multibanding'' techniques~\cite{Vinciguerra:2017ngf}.
This model can be used to determine the impact on \ac{GW} searches,
event parameter estimation and
population inference due to the effects of precession and higher modes.

We expect to be able to greatly improve {\tt PhenomPv3HM}, and similar models, by using
models for the underlying higher multipole aligned-spin model that have
been calibrated to \ac{NR} waveforms~\cite{cecilioplus}. Likewise, a model for the precession dynamics
tuned to precessing \ac{NR} simulations will improve its performance~\cite{hamiltonplus}.
Although our model is a function of the 7 dimensional intrinsic parameter space of
non-eccentric \ac{BBH} mergers it is not 7 dimensional across the entire
coalescence. It is true during the inspiral, but during the pre-merger
and merger we use an effective aligned-spin parametrization.
Work is underway to develop an \ac{NR} calibrated aligned-spin model with the
effects
of two independent aligned-spins~\cite{geraintplus}. In addition, promising
attempts to dynamically enhance incomplete models via
singular-value-decomposition have recently been
presented~\cite{Setyawati:2018tqp}, and the model introduced here can easily
be employed by such an automated tuning process.

With regards to higher modes we only include a subset of the complete
list of modes, specifically $(\ell, |m|) = ((2,2), (2,1), (3,3), (3,2), (4,4), (4,3))$.
We also ignore mode mixing~\cite{Berti:2014fga}
and the asymmetry between the $+m$ and $-m$ modes, which are
responsible for out of plane recoils~\cite{Brugmann:2007zj}.

We plan to extend this model to include tidal effects
as introduced in~\cite{Dietrich:2018uni,Dietrich:2019kaq} as well as implement
a model for the \ac{GW} suitable for neutron star-black hole binaries
where the effects of spin-precession, subdominant multipoles and tidal effects
could all become important. The model presented here could be
used as a baseline for such a model.
Finally this model is implemented in the
LSC Algorithm Library (LAL)~\cite{lalsuite}.

\begin{acknowledgments}

We thank Geoffrey Lovelace, Lawrence Kidder and Michael Boyle for support
with the SXS catalogue.
We thank Carl J. Haster and Yoshinta Setyawati for useful discussions.
S.K. and F.O. acknowledge support by the
Max Planck Society’s Independent Research Group Grant.
The Flatiron Institute is supported by the Simons Foundation.
M.H. was supported by Science and Technology Facilities Council (STFC) grant ST/L000962/1
and European Research Council Consolidator Grant 647839, and thanks the Amaldi Research
Center for hospitality.
We thank the Atlas cluster computing team at AEI Hannover.
The authors are grateful for computational resources provided by
the LIGO Laboratory and supported by National Science Foundation Grants
PHY-0757058 and PHY-0823459.
This work made use of numerous open source computational packages
such as python~\cite{python}, NumPy, SciPy~\cite{scipy}, Matplotlib~\cite{Hunter:2007},
the \ac{GW} data analysis software library {\tt pycbc}~\cite{pycbc-software}
and the LSC Algorithm Library (LAL)~\cite{lalsuite}.
This research has made use of data, software and/or web tools obtained from the Gravitational Wave Open Science Center (https://www.gw-openscience.org), a service of LIGO Laboratory, the LIGO Scientific Collaboration and the Virgo Collaboration. LIGO is funded by the U.S. National Science Foundation. Virgo is funded by the French Centre National de Recherche Scientifique (CNRS), the Italian Istituto Nazionale della Fisica Nucleare (INFN) and the Dutch Nikhef, with contributions by Polish and Hungarian institutes.

\end{acknowledgments}

\bibliography{phenompv3hm}

\end{document}